\documentclass[a4paper,11pt]{article}
\pdfoutput=1 
\usepackage{jheppub} 
\usepackage[T1]{fontenc}
\usepackage{amsmath}
\usepackage{xcolor}
\usepackage{braket}
\usepackage{comment}
\usepackage{hyperref}
\usepackage{amsfonts}
\usepackage{ulem}
\usepackage{mathtools}
\usepackage{dsfont}
\usepackage[utf8]{inputenc}
\usepackage{bbm}
\usepackage{caption}
\usepackage{subcaption}

\usepackage{simpler-wick}
\usepackage{simplewick}

\usepackage{graphicx}
\graphicspath{{./Figures/}}

\newcommand{\cO}{\mathcal{O}}

\newcommand{\cH}{\mathcal{H}}

\newcommand{\bea}{\begin{eqnarray}}
\newcommand{\eea}{\end{eqnarray}}
\newcommand{\be}{\begin{equation}}
\newcommand{\ee}{\end{equation}}
\newcommand{\bne}{\begin{equation}}
\newcommand{\ene}{\end{equation}}

\def\bne{\begin{equation}}
\def\ene{\end{equation}}

\newcommand{\overbar}[1]{\mkern 1.5mu\overline{\mkern-1.5mu#1\mkern-1.5mu}\mkern 1.5mu}

\newcommand{\Wg}{\mathsf{Wg}}
\newcommand{\Tr}{\operatorname{Tr}}

\title{\boldmath 

Page curves and replica wormholes from random dynamics}

\author[a]{Jan de Boer,}
\author[a]{Jildou Hollander,}
\author[a,b]{and Andrew Rolph}

\affiliation[a]{Institute for Theoretical Physics, University of Amsterdam, PO Box 94485, 1090 GL Amsterdam, The Netherlands}
\affiliation[b]{Vrije Universiteit Brussel (VUB), Pleinlaan 2, B-1050 Brussels, Belgium}

\emailAdd{j.deboer@uva.nl}
\emailAdd{j.s.hollander@uva.nl}
\emailAdd{andrew.d.rolph@gmail.com}

\abstract{
We show how to capture both the non-unitary Page curve and replica wormhole-like contributions that restore unitarity in a toy quantum system with random dynamics. The motivation is to find the simplest dynamical model that captures this aspect of gravitational physics. 
In our model, we evolve with an ensemble of Hamiltonians with GUE statistics within microcanonical windows. The entropy of the averaged state gives the non-unitary curve, the averaged entropy gives the unitary curve, and the difference comes from matrix index contractions in the Haar averaging that connect the density matrices in a replica wormhole-like manner. 

}

\begin{document} 
\maketitle
\flushbottom

\section{Introduction}
\label{sec: introduction}

The black hole information problem is a tension between unitarity in quantum gravity and the insensitivity of Hawking radiation to initial conditions~\cite{Hawking1975,Hawking1976}.
Pure state black holes evaporate away to mixed-state thermal radiation. 
The von Neumann entropy of black hole radiation is sensitive to this tension because Hawking radiation is thermal and evaporation implies that the entropy increases monotonically, while unitarity requires it to return to zero~\cite{Page1993information,Page1993average,Page2013}. A resolution to the problem, as far as calculating a unitary Page curve, recognises that the exact state and the semiclassical state prepared by the dominant gravitational saddle are not the same, and in particular, their entropies can be quite different. This was first seen in holographic systems through the quantum extremal surface formula~\cite{Penington2020, Almheiri2019entropy, Almheiri2020page}, which can calculate the entropy of the exact state in the boundary CFT~\cite{Ryu2006, Hubeny2007,Faulkner2013,Engelhardt2015,Rolph:2021hgz}, then generalised to non-holographic systems with the island formula which can be derived by including novel connected gravitational saddles in the replica trick called replica wormholes~\cite{Almheiri2020replica,Penington2022replica}.

In a way, AdS/CFT itself is a resolution to the black hole information paradox. For small AdS black holes, the information paradox is the prediction that a pure state black hole will evolve to a mixed state thermal gas, and AdS/CFT is a resolution in the sense that the dual description of evaporation is manifestly unitary, so the von Neumann entropy is simply zero at all times.\footnote{The paradox is cleanest in this set-up, but we can also couple the boundary CFT to a radiation bath to get the standard non-trivial Page curve.} This resolution is unsatisfying, but why? This is because it is not clear how unitarity is restored in the bulk. In principle, to know everything about AdS black hole evaporation, one can simply take a high energy CFT state, let it evolve, and then probe it. In practice, the state is enormously complex and it is impossible to semiclassically determine microscopic details, for example, how the exact dynamics of the late time radiation purifies the early radiation. The same objection holds for the island formula: it gives us unitary Page curves, but it is not obvious how the Euclidean gravitational path integral from which it is derived tells us anything about exactly how the late time radiation purifies the early radiation. 
 
To make progress towards this fuller resolution, let's ask: what, for an evaporating AdS black hole, are the necessary and sufficient conditions on the holographic dual to get a unitary Page curve?\footnote{The Page curve in this context is the renormalised von Neumann entropy of the whole boundary CFT for a small isolated AdS black hole, or of the bath if the holographic CFT is coupled to one.}
Requiring exact unitarity in a quantum system with Hilbert space dimension $|\cH|$ imposes $2|\cH|^2$ equality constraints on the matrix elements of the Hamiltonian, whereas a unitary Page curve is a single curve, so exact unitary is presumably overkill. 
Does only a subsector of the theory need to be unitary and, if so, which?
Can we replace unitary time evolution with a time-dependent completely positive and trace-preserving (CPTP) map and still get a unitary Page curve?
Can exact properties of the Hamiltonian be replaced with statistical properties and, if so, which? For this last point, there is a hint that this is true, because (1) replica wormholes give unitary Page curves, (2) replica wormholes are gravitational saddles, and (3) AdS gravity, without stringy corrections, is conjectured to be best described by a statistical ensemble (along the lines developed in \cite{Belin2021,Belin2022generalized,Belin2022non,Anous2022ope,Belin2023approximate,ignorance}, see also \cite{Pollack2020eigenstate}). Answering these questions will tell us which parts of the non-gravitational dual are responsible for the late time purification of Hawking radiation, and which are irrelevant. 

In this paper, to deepen our understanding of the underlying mechanics of unitary black hole evaporation, we show how to capture both the non-unitary Page curve and replica wormhole-like contributions that restore unitarity in a toy quantum system with random dynamics.
Semiclassical gravity gives us two curves for the radiation entanglement entropy: the Hawking curve and, after including replica wormholes, the unitary Page curve. Our claim is that chaotic dynamics is, with a few other basic assumptions, sufficient to capture both. If semiclassical gravity captures only the statistical properties of the exact UV theory, then those statistical properties alone should be sufficient. As far as these two curves are concerned, neither complex gravitational systems nor holography are needed. 

Our model has a factorised Hilbert space of the system $\cH_{\mathsf{A}}$ and its environment $\cH_{\mathsf{B}}$. One can imagine the system representing a gravitational system with a black hole, or a lab with a piece of coal. Within the full spectrum, we consider a high-energy microcanonical window, which we take to spanned by $N$ product states, one of which has the energy in the system, so that the environment is in its vacuum state,
\bne
\Tr_{\mathsf{A}} (\ket{\psi_1}\bra{\psi_1}) = \ket{0} \bra{0}_{\mathsf{B}},
\ene 
The other $(N-1)$ states have all the energy in the environment and none in the system: 
\bne
\Tr_{\mathsf{B}} (\ket{\psi_2}\bra{\psi_2}) = \dots = \Tr_{\mathsf{B}} (\ket{\psi_N}\bra{\psi_N}) = \ket{0}\bra{0}_{\mathsf{A}}.
\ene 
These represent the radiation states and they are entropically favoured within the microcanonical ensemble.
We take the initial state to be the special state $\ket{\psi_1}$, with the energy in the system, and this state represents the black hole or piece of coal. 

A bit more informally, our Hilbert space therefore looks like
\bne
{\cal H}= \big(|{\rm black\, hole}\rangle_{\mathsf{A}}\otimes |{\rm empty}\rangle_{\mathsf{B}}\big) \, \oplus \,\big( |{\rm empty}\rangle_{\mathsf{A}} \otimes {\cal H}_{\rm rad,\mathsf{B}}\big),
\ene
with ${\cal H}_{\rm rad,\mathsf{B}}$ the $(N-1)$-dimensional space of radiation states in the $\mathsf{B}$-system.

\begin{figure}
    \centering
    \includegraphics[width=\textwidth]{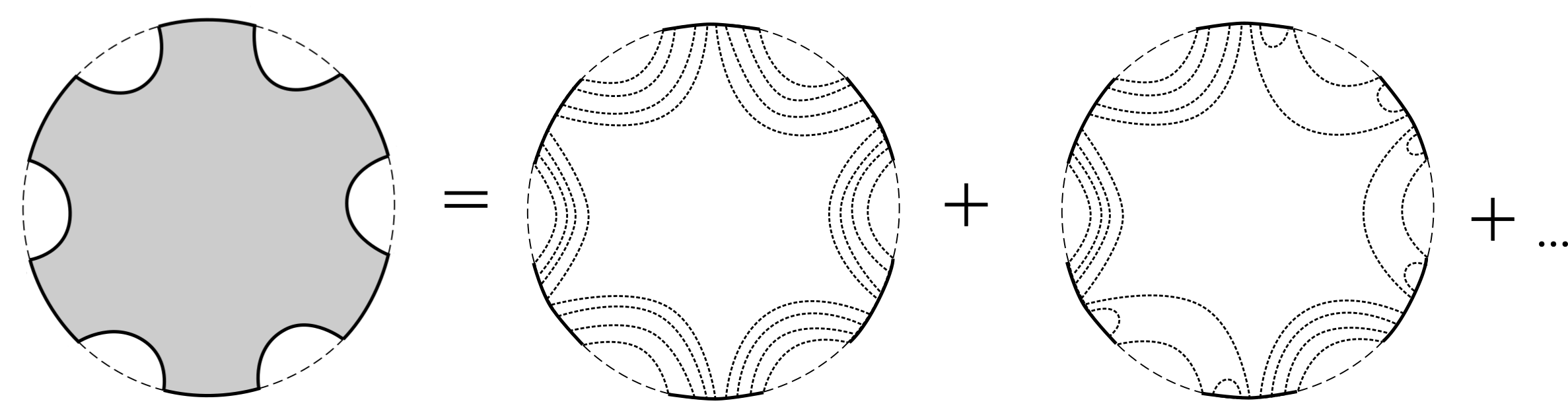}
    \caption{In our model, replica wormholes, in the sense of contributions to the R\'enyi entropies that connect the density matrices and restore unitarity, come from chaotic dynamics through the Haar averaging. The density matrices are quartic functions of the Hamiltonian-diagonalising unitary matrix, and taking the Haar average of the R\'enyi entropy is what connects the density matrices. The dashed lines represent unitary matrix index contractions. }
    \label{fig: sum of contractions}
\end{figure}

The matrix elements of the Hamiltonian within this narrow energy band, $H_{ij} = \bra{\psi_i}H \ket{\psi_j}$, are taken to be random with GUE matrix statistics, with probability measure
\bne \mu(H_{ij}) \propto \exp\left ( -\frac{N}{2}|H_{ij}|^2 \right). \ene
Evolution with this random Hamiltonian generates a time-dependent distribution of pure states. 
For a given draw of a random Hamiltonian, the reduced density matrices of the system and environment, and their R\'enyi entropies, are functions of the Hamiltonian's diagonalising unitary matrix and its eigenvalues. 
Averaging over the unitary matrices, with the Haar measure, connects matrix indices. 
The R\'enyi entropies have contributions that connect the density matrices and disconnected components.
In Fig.~\ref{fig: sum of contractions} we diagrammatically represent some of the matrix contractions from the Haar averaging. 
As for replica wormholes, the contributions that restore unitarity at late times connect the density matrices.

The result for the $n$'th R\'enyi entropy of the averaged environment density matrix $\rho_{\mathsf{B}}$, in the large $N$ approximation, is
\bne S^{(n)} \left(\overline{\rho_{\mathsf{B}}}(t)\right)= 
\frac{1}{1-n}\log \left( \frac{1}{N^{n-1}}\left(1 - \overline{g}(t)\right)^n+\overline{g}^n(t)\right),
\ene
where $\overline{g}(t)$ is the averaged normalised spectral form factor, while the averaged R\'enyi entropy is\footnote{Here we take the average before taking the logarithm. We thus compute the annealed instead of the quenched averaged R\'enyi entropies.}
\footnote{We implicitly assume that $\overline{g^n}(t)=\overline{g}(t)^n$. We will come back to the validity of this assumption, but expect it to hold for time scales relevant to us.}
\bne \overline{S^{(n)} (\rho_{\mathsf{B}}(t))} = \frac{1}{1-n}\log \left(\left(1 - \overline{g}(t)\right)^n + \overline{g}^n(t)\right).
\ene
\begin{figure}
    \centering
    \includegraphics[width=0.7\textwidth]{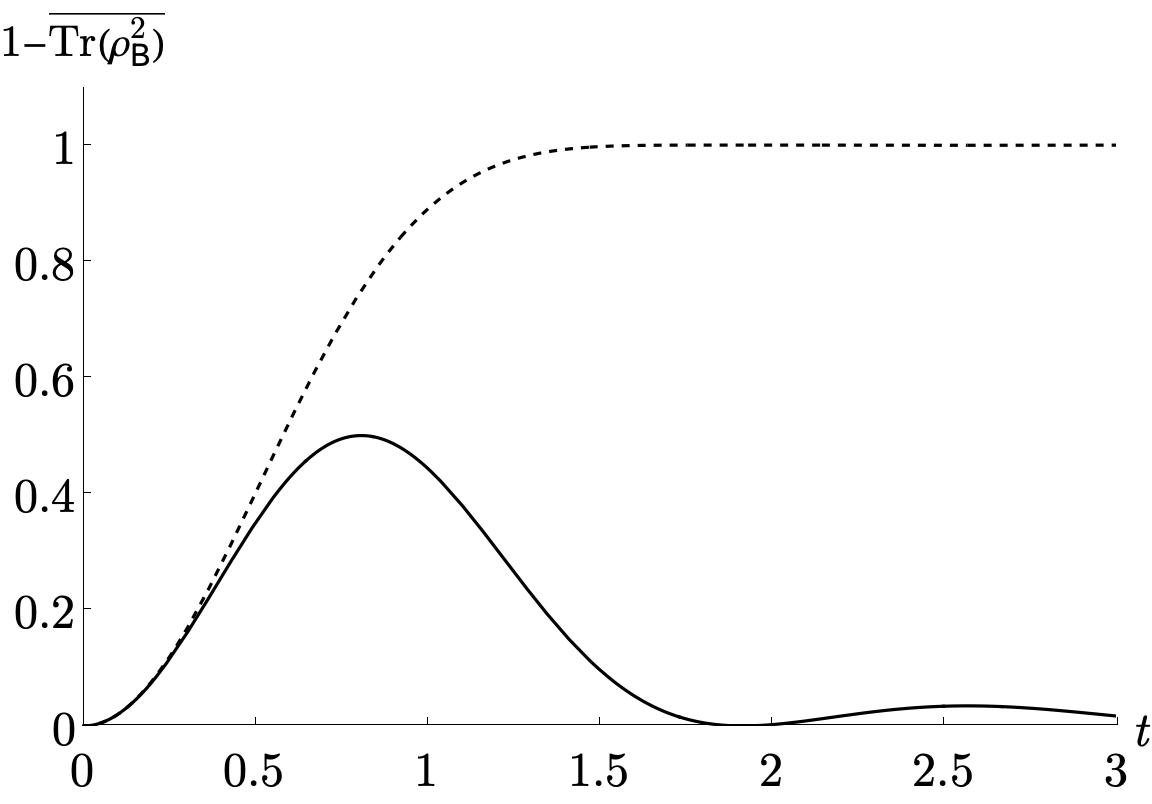}
    \caption{
    The averaged purity of the radiation in our model with GUE-random Hamiltonians.
    The dashed line is the result that neglects connected replica wormhole-like contributions from the Haar averaging, and shows non-unitary evolution, while the solid line does take into account these connected contributions, and shows unitary evolution.
    }
    \label{fig: Page analogue}
\end{figure}
In Fig.~\ref{fig: Page analogue} we plot the purity of the radiation density matrix.
We calculate the von Neumann entropies by analytically continuing our results for the R\'enyi entropies and we find both the unitary and non-unitary Page curves. 

While there are many features of black hole evaporation that our simple model does not capture, our goal is only to capture one particular aspect: the qualitative shape of the unitary and non-unitary Page curves.
Our model captures this, and the difference between the curves comes through the Haar averaging of the random dynamics from the subset of matrix index contractions that connects the density matrices in a replica wormhole-like way. 

It is perhaps worth emphasizing two key aspects of our model. The first is that one needs to have a mechanism which guarantees that most of the information will be transferred to the radiation system $\mathsf{B}$. In many examples including our toy model, this follows from fairly straightforward entropy and energy considerations as we review in Sec.~\ref{sec: toy model}. The second key aspect is that the random dynamics maps an initial pure state to a {\it classical statistical} mixture of pure states at later times. Computations which involve only one replica of the system cannot distinguish a classical statistical mixture of pure states from a corresponding single mixed state as both yield the same expectation values. However, replica computations can tell the difference between the two. In particular, the expectation value of the purity is one for classical statistical mixtures of pure states, while it is smaller than one for mixed states. The ability of replica computations to distinguish classical mixtures of pure states from mixed states is what allows one to recover a Page curve using only semi-classical computations. Note that this does not imply that semi-classical computations predict the precise final state, for this one would need the actual UV completion of the system.

Alternatively, one can also view the success of semi-classical replica computations in producing the correct Page curve as a confirmation of the validity of the statistical interpretation of semi-classical gravity.

The plan for the remainder of this paper is as follows.
In Sec. \ref{sec: toy model}, we detail the set-up of our toy model of black hole evaporation, and calculate the density matrices, the R\'enyi entropies, and von Neumann entropies, with and without averaging with respect to the random Hamiltonian. 
In Sec. \ref{sec: Extensions}, we discuss the features of black hole evaporation that our model does not capture, how the model could be modified and extended, and further applications.
In Sec. \ref{sec: conclusion and discussion}, we discuss the results and conceptual takeaways and give possible directions for future research.

\section{Toy model of black hole evaporation}
\label{sec: toy model}
After a black hole has evaporated, or a piece of coal burned away, we expect the final state of the radiation to be pure. Why? To have a model where the unitary Page curve returns to zero, we need to understand this.

Suppose we have a system $\mathsf{A}$ coupled to its environment $\mathsf{B}$. At late times, for the environment's reduced state to be pure, it is sufficient if~\cite{Marolf:2017jkr}:
\begin{enumerate}
\item The final state of the system is a ground state.
\item The ground state of the system is non-degenerate.
\item The evolution of the system and environment is unitary.
\item The initial state of the system and environment is pure.
\end{enumerate}
Conditions (1) and (2) imply that the system's final state is pure. Conditions (3) and (4) imply that the final state of the system and environment is pure. Together they imply that the final state of the environment is pure.

Let us assume conditions (2)-(4). For a given initial state, condition (1) is satisfied if energy is conserved and all of it leaves the system to the environment. Whether this happens depends on the dynamics, and whether microstates where the energy has left the system are entropically favoured. It also requires the system's spectrum to be gapped, because if there is a continuous spectrum above the ground state for example, then the energy of the system can be arbitrarily low, while the entropy could be arbitrarily high. 
Assuming energy conservation and ergodic evolution, the long-time average of the energy in $\mathsf{A}$ will equal the ensemble average. If nearly all the microstates in the ensemble have zero energy in $\mathsf{A}$, then the ensemble average of the energy 
in $\mathsf{A}$ will be small. If the ensemble average is much smaller than the gap between the (non-degenerate) ground state and the first excited state in $\mathsf{A}$, then the long-time average of the reduced density matrix on $\mathsf{A}$ will be very close to the pure ground state.

A somewhat more quantitative description is as follows. Suppose we have a pure state $|\psi\rangle$ in ${\cal H}_\mathsf{A}\otimes {\cal H}_\mathsf{B}$ where the Hilbert spaces have dimensions $d_\mathsf{A}$ and $d_\mathsf{B}$ with $d_\mathsf{A}\ll d_\mathsf{B}$. The entanglement entropy between $\mathsf{A}$ and $\mathsf{B}$ is then bounded by $\log d_\mathsf{A}$. Now suppose that there is an operator $H_\mathsf{A}$ acting on the $\mathsf{A}$-system with lowest eigenvalue $E_0$ and next lowest eigenvalue $E_0+E_{\rm gap}$. If the expectation value of $H_\mathsf{A}$ in the state $|\psi\rangle$ (i.e. ${\rm Tr}(\rho_\mathsf{A} H_\mathsf{A})$) equals $E$ and $\epsilon=(E-E_0)/E_{\rm gap}$ is much smaller than one, then the entanglement entropy for the state $|\psi\rangle$ is bounded by
\bne
S_\mathsf{A} \leq -(1-\epsilon)\log(1-\epsilon) - \epsilon\log\epsilon + \epsilon \log (d_\mathsf{A}-1).
\ene
This is much smaller than one as long as $\epsilon \ll 1/\log(d_\mathsf{A}-1)$ and the reduced density matrix is close to that of a pure state. If we view $E$ as the energy in the $\mathsf{A}$-system and we equipartition the total energy $E_{\rm tot}$ of an initial state over the $\mathsf{A}$ and $\mathsf{B}$-system then the condition $\epsilon \ll 1/\log(d_\mathsf{A}-1)$ becomes
\bne
\frac{ \frac{d_\mathsf{A}}{d_\mathsf{A}+d_\mathsf{B}} E_{\rm tot} - E_0}{E_{\rm gap}}\ll \frac{1}{\log(d_\mathsf{A}-1)}\, .
\ene
which can always be achieved by making $d_\mathsf{B}/d_\mathsf{A}$ sufficiently large.

The takeaway is that we want our model to have energy conservation, ergodicity, a gapped spectrum, and entropic dominance of microstates where the energy has left the system.

\subsection{Set-up}
Here we describe the set-up for our toy model.
We want the model to be as simple as possible, but still capable of dynamically capturing the following semiclassical gravitational predictions qualitatively: (1) the mixed Hawking radiation state and the non-unitary Page curve, (2) replica wormholes, and
(3) the unitary Page curve.

First, we describe the structure of the Hilbert space and the space of states available under time evolution. The full Hilbert space $\cH$ is a tensor product of system and environment Hilbert spaces: 
\bne \cH = \cH_{\mathsf{A}} \otimes \cH_{\mathsf{B}}.\ene 
In general, we can decompose Hilbert spaces into direct sums over mutually orthogonal eigenspaces of any compact self-adjoint operator, by the spectral theorem. Let us decompose our $\cH$ into a direct sum of microcanonical Hilbert subspaces with energies binned into narrow energy windows: 
\bne \cH = \bigoplus_E \cH_{\text{micro};E\pm \delta E}. \ene
Each $\cH_{\text{micro};E\pm \delta E}$ is a microcanonical Hilbert subspace spanned by the energy eigenstates with energies in the range $[E-\delta E, E+ \delta E]$. 

We are decomposing the Hilbert space this way to allow us to account for energy conservation and because it is a reasonable assumption that our initial state, if it represents a black hole microstate, has support only within a high energy microcanonical subspace $\cH_{\text{micro};E\pm \delta E}$. Since time evolution cannot take the states out of this subspace, if energy is conserved, only this subspace is important. This is a feature of our model that differentiates it from those which treat black hole evaporation as a random unitary on $\cH_\mathsf{A} \otimes \cH_\mathsf{B}$~\cite{Page1993information, Page1993average}, which does not conserve energy.

If the system and environment are weakly interacting, then product states $\ket{E-E'}_{\mathsf{A}} \otimes \ket{E'}_{\mathsf{B}}$, which are eigenstates of the decoupled Hamiltonian with energy $E$, will be approximate eigenstates of the coupled Hamiltonian, meaning that they are superpositions of states in $\cH_{\text{micro};E\pm \delta E}$. Then, it is a reasonable approximation that $\cH_{\text{micro};E\pm \delta E}$ is also spanned by product states like $\ket{E-E'}_{\mathsf{A}} \otimes \ket{E'}_{\mathsf{B}}$, so that
\bne \cH_{\text{micro};E \pm \delta E} \simeq \bigoplus_{E'} \left(\cH_{\mathsf{A};E-E'} \otimes \cH_{\mathsf{B};E'} \right). \label{eq:micro1}\ene
A state in $\cH_{\mathsf{A};E-E'} \otimes \cH_{\mathsf{B};E'}$ is interpreted in our model as a partially evaporated black hole, or burning piece of coal, that had initial energy $E$ and has lost energy $E'$ to the environment. To give a reasonable ansatz for the sizes of the Hilbert spaces, for a small evaporating $\text{AdS}_4$ black hole, $\log |\cH_{\mathsf{A};E}| = S_{\text{BH}}(E) \sim (E l_p)^2$ and $\log |\cH_{\mathsf{B};E}| = S_{\text{rad}} (E) \sim (E l_{\text{AdS}})^{3/4}$. 

Recall that for the system's final state to be pure, the essential ingredients are: no ground state degeneracy, a gap in the spectrum, and entropic dominance of microstates where all the energy is in the environment. To that end, and to further simplify the space of states that are energetically accessible, we assume that there is one black hole microstate in $\cH_{\text{micro};E\pm \delta E}$, and $N-1$ energetically accessible radiation states: 

\begin{equation} \cH_{\text{micro};E\pm \delta E} = 
(\ket{E}_{\mathsf{A}} \otimes \ket{0}_{\mathsf{B}} ) \oplus 
(\ket{0}_{\mathsf{A}} \otimes \cH_{\mathsf{B};E} ).
\label{eq: micro Hilbert}\end{equation}
We take the microcanonical entropy to be large,
\bne e^{S_{\text{micro}}(E)}:= |\cH_{\text{micro};E\pm \delta E}|  = N \gg 1, \ene 
and all but one of the microstates have the energy in the environment, $|\cH_{\mathsf{B};E}| = N-1$. 

Our goal is to capture the qualitative rather than quantitative features of radiation density matrices and Page curves in the simplest toy model. While~\eqref{eq: micro Hilbert} has no partially-evaporated states and only one black hole microstate, we will show that it is sufficient to capture the qualitative features. 

We choose an orthonormal basis $\ket{i}_{\mathsf{B}}$ for states in $\cH_{\mathsf{B};E}$, with $i = 2,\dots, N$, which in turn fixes an orthonormal basis for $\cH_{\text{micro};E\pm \delta E}$, 
\bne \ket{\psi_i} :=
\begin{cases}
    \ket{E}_\mathsf{A} \ket{0}_\mathsf{B}  &\text{ for $i=1$,}\\
    \ket{0}_\mathsf{A} \ket{i}_\mathsf{B} &\text{ for $i=2,\dots,N$.}
    \label{eq: states}
\end{cases} 
\ene
with $\braket{\psi_i | \psi_{j}} = \delta_{i,j}$.

Our initial state is $\ket{\psi_1}$. It is a non-stationary state within a particular microcanonical window $\cH_{\text{micro};E\pm \delta E}$. It is an atypical state because all the energy is within the $\mathsf{A}$ subsystem, and its time evolution is towards typical microstates where all the energy is within the environment. 

Only the projection of the Hamiltonian onto the microcanonical window is important for the time evolution of $\ket{\psi_1} \in \cH_{\text{micro}, E\pm\delta E}$. 
By energy conservation, time evolution cannot take us out of this energy window, which implies that the projected Hamiltonian satisfies
\bne H' := P H P = H P, \ene
where $P$ is the projection operator

\bne P := \sum_{i=1}^N \ket{\psi_i}\bra{\psi_i}. \ene
 As claimed, the time evolution only depends on the projected Hamiltonian: 
\begin{equation}
    \rho (t) := e^{iHt}\ket{\psi_1}\bra{\psi_1}e^{-iHt} = e^{iH't}\ket{\psi_1}\bra{\psi_1}e^{-iH't}.
    \label{eq: H vs Hint}
\end{equation}
$H'$ still has a large diagonal component, $H' \approx E \cdot \mathds{1}$, which we subtract off as it does not affect $\rho(t)$. 
The off-diagonal matrix elements of $H'$ come from the coupling between $\mathsf{A}$ and $\mathsf{B}$ and vanish if the coupling is turned off. For small but finite coupling, for a many-body system, we cannot generally solve to find the exact eigenvalues of $H'$ or the unitary matrix that diagonalises $\bra{\psi_i}H' \ket{\psi_j}$. We do however expect $\bra{\psi_i}H' \ket{\psi_j}$ to look like a $N\times N$ random matrix and its statistics to be predicted by random matrix theory (RMT). 

 Chaotic quantum many-body systems exhibit a strong form of universality \cite{Berry1985semiclassical, Altland1997nonstandard, Deutsch1991quantum,Srednicki1994chaos,DAlessio:2015qtq}. 
RMT captures certain universal aspects of chaotic systems, such as the statistics of energy levels and matrix elements of observables \cite{Wigner1955characteristic,Wigner1957characteristic,Wigner1958distribution,Dyson1962statistical}.  The spectral statistics within sufficiently narrow energy bands of quantum systems with classically chaotic counterparts are believed to be captured by RMT~\cite{Bohigas1984characterization}. This assumption of RMT behaviour within narrow energy bands is a weaker assumption than ETH~\cite{DAlessio:2015qtq}.  
 High energy eigenstates in chaotic theories are of the form $\sum M_{ab} \ket{E_a}\otimes\ket{E_b}$ 
with $M_{ab}$ a banded random matrix, with bandwidth controlled by the strength of the interaction~\cite{Deutsch_2010,Murthy:2019qvb}. 

In our model, we replace time evolution with a single projected Hamiltonian $H'$ by an ensemble of Hamiltonians with the same statistical properties as $H'$, in particular, the symmetry class and the spectral distribution. In what follows, we will always assume that this ensemble is unitarily invariant, as this symmetry class is the least restrictive on the Hamiltonian.

When needed, for tractability and for explicit results, we will further assume that the ensemble is Gaussian. Gaussian ensembles have probability measures
\begin{equation}
\mu(H) \propto \exp \left(-\frac{1}{a^2 } \Tr H^2\right),
\label{eq: Gaussian ensemble measure}
\end{equation}
where $a$ sets the overall energy scale.
When we specify ensembles of Hamiltonians $H$, like above, we will always be implicitly referring only to the distribution of the relevant matrix elements $\bra{\psi_i}H \ket{\psi_j}$, which are the same as the projected Hamiltonian $H'$. Without loss of generality, we also choose ensembles whose mean is zero because the diagonal piece of $H' \approx E \cdot \mathbbm{1}$ does not affect $\rho(t)$.

To recap our model, we have an initial state $\ket{\psi_1}$, and we are evolving with a unitary-invariant ensemble of $N\times N$ Hamiltonian matrices $\bra{\psi_i}H\ket{\psi_j}$, with $\ket{\psi_i}$ defined in~\eqref{eq: states}.
We will show that the ensemble average of whatever quantity we are considering, density matrices or R\'enyi entropies, qualitatively equals the semiclassical gravitational approximation to the same. We will calculate reduced density matrices and entanglement measures before and after averaging, and connect the results to the unitary and non-unitary Page curves and replica wormholes. Evolution with an ensemble of Hamiltonians is fundamentally how non-unitarity is introduced into the model.

\subsection{Exact density matrices}
\label{sec: Density matrices}
We will first calculate the density matrices, time-evolved with a single random $H$ drawn from the ensemble. 
The Hamiltonian can be diagonalised to $H = U \Lambda U^\dagger$ and, since $H$ is Hermitian, $U$ is unitary and $\Lambda$ is diagonal with real eigenvalues. When expressed in the $\ket{\psi_i}$ basis, the Hamiltonian is 
\begin{equation}
    \bra{\psi_j}H \ket{\psi_l} = \sum_{k} \lambda_k U_{jk} U^\dagger_{kl}.
\end{equation}
The probability distributions of the eigenvalues $\lambda$ and unitary matrices $U$ depend on the ensemble, and later we will specialise to the GUE.

The time-dependent density matrix can be found by acting with the time evolution operator $e^{iHt}$ on the initial state, such that
\begin{equation}
\begin{split}
     \rho(t)
    &= \sum_{i,j,k,l=1}^N  U_{ik}U_{k1}^\dagger U_{1l}U_{lj}^\dagger ~ e^{i(\lambda_k-\lambda_l)t}\ket{\psi_i}\bra{\psi_j}.
    \label{eq: rho(t)}
\end{split}
\end{equation}
Recall that the microcanonical Hilbert space \eqref{eq: micro Hilbert} is embedded in the factorizable Hilbert space $\cH = \cH_{\mathsf{A}}\otimes \cH_{\mathsf{B}}$, so partial traces are defined within this larger Hilbert space. 
Tracing out subsystem ${\mathsf{B}}$, the only non-vanishing terms have $i=j$. The partial density matrix $\rho_{\mathsf{A}}(t)$ is thus given by
\begin{equation}
    \begin{split}
        \rho_{\mathsf{A}}(t) 
         &= \sum_{k,l=1}^N e^{i(\lambda_{k}-\lambda_{l})t}  \left( U_{1k}U_{k1}^\dagger U_{1l}U_{l1}^\dagger\ket{E}\bra{E}_{\mathsf{A}}+\sum_{i=2}^N U_{ik}U_{k1}^\dagger U_{1l}U_{li}^\dagger \ket{0}\bra{0}_{\mathsf{A}}\right).
    \end{split}
    \label{eq:rhoAt}
\end{equation}
This $2\times 2$ density matrix is diagonal. If we trace out subsystem ${\mathsf{A}}$ instead, the partial density matrix of subsystem ${\mathsf{B}}$ is given by
\begin{equation}
    \rho_{\mathsf{B}}(t) 
        = \sum_{k,l=1}^Ne^{i(\lambda_{k}-\lambda_{l})t} \left( U_{1k}U_{k1}^\dagger U_{1l}U_{l1}^\dagger\ket{0}\bra{0}_{\mathsf{B}}+\sum_{i,j =2}^N U_{ik}U_{k1}^\dagger U_{1l}U_{lj}^\dagger \ket{i}\bra{j}_{\mathsf{B}}\right).
    \label{eq: rhoB}
\end{equation}
Unlike $\rho_{\mathsf{A}}(t)$, the reduced density matrix in \eqref{eq: rhoB} is not diagonal.

\subsection{Averaged density matrices}
\label{sec: Average states}
Next, we calculate the averaged density matrices and their purity and R\'enyi entropies. Our averaging is over random Hamiltonians with an appropriate probability measure: $\overline{f(H)} := \int d\mu (H) f(H)$. We will soon restrict the computation to the GUE measure, but for now let the measure be any that is invariant under unitary transformations: $\mu (H) = \mu( U H U^\dagger)$. 

One approach to calculating averaged R\'enyi entropyies is to first calculate the joint probability distribution function (PDF) of the entanglement spectra of the reduced density matrices. For us, since $\rho_\mathsf{A}$ is a $2\times 2$ matrix, and $\Tr(\rho_\mathsf{A}) =1$, there is only one independent eigenvalue in the entanglement spectrum between $\mathsf{A}$ and $\mathsf{B}$. Nonetheless, we were unable to determine that PDF, so we have relegated the results to App.~\ref{app:spectrumPDF}. 

Our approach will be to average the density matrices over $H$: we first integrate over the diagonalising unitaries, then the eigenvalues. See App.~\ref{app:weingarten} for details on how to perform unitary integrals.
The averaged reduced density matrix for $\mathsf{A}$ is
\begin{equation}
    \overline{\rho_{\mathsf{A}}}(t)
     = \frac{N+\overline{W}(t)}{N(N+1)}\ket{E}\bra{E}_{\mathsf{A}} + \frac{N^2-\overline{W}(t)}{N(N+1)}\ket{0}\bra{0}_{\mathsf{A}}.
     \label{eq:rhoAbar}
\end{equation}
We have introduced the spectral form factor:
\begin{equation} \begin{split}
    W(t) := \Tr\left(e^{iHt}\right)\Tr \left(e^{-iHt}\right) 
     =\sum_{k,l=1}^N e^{i(\lambda_{k}-\lambda_{l})t} .
\end{split}\end{equation}
Most of our results only require unitary invariance of the measure. The only place where the actual details of the measure come into play, are in the exact time dependence of this spectral form factor, which depends on the ensemble under consideration. Even without restricting to a specific ensemble, chaotic systems have some universal behavior of the spectral form factor.
From its definition, $W(t=0) = N^2$ and, in chaotic systems, the long-time average of $W$ is order $N$.\footnote{The long time average of $\Tr(e^{iHt})$ is a sum of random phases from the uniform distribution, so $W$ is the square of the absolute value of an $N$-step random walk in the complex plane.} 
To elucidate the leading order terms, we introduce the variable $g(t)$, defined as the ratio of $W(t)$ to $N^2$:
\begin{equation}
    g(t) \coloneqq \frac{W(t)}{N^2}~.
\end{equation}
At early times, $g(t)$ is of order one. However, as time progresses, it decays to a value of $1/N$ during late-time evolution. 
The spectral form factor is only a function of the Hamiltonian's eigenvalues and time and the precise function for $g(t)$ depends on the eigenvalue probability distribution of the ensemble. 
In what follows, we will sometimes omit the explicit $t$-dependence and just write $g$. 
The von Neumann entropy of the averaged density matrix in the large $N$ approximation is 
\begin{equation} \label{eq:SrhoAaveg}
    S(\overline{\rho_{\mathsf{A}}}(t)) = -\overline{g}(t) \log \left(\overline{g}(t)\right)- (1-\overline{g}(t)) \log \left(1-\overline{g}(t)\right).
\end{equation}

Thus, in the large $N$ approximation, as depicted in Fig.~\ref{fig:naive entropy} for the GUE, the entropy~\eqref{eq:SrhoAaveg} starts at zero, increases to its maximum value $S_{\max}=\log 2$, and subsequently decreases to approximately zero at late times. 

 Let us now look at the behaviour of the von Neumann entropy of the average state in the environment, subsystem ${\mathsf{B}}$. 
 The averaged partial density matrix $\overline{\rho_{\mathsf{B}}}(t)$ is given by
\begin{equation}
    \begin{split}
        \overline{\rho_{\mathsf{B}}} (t)
        &= \frac{1+\overline{g}(t)N}{N+1}\ket{0}\bra{0}_{\mathsf{B}} + \frac{N-\overline{g}(t)N}{N^2-1}\sum_{i =2}^N \ket{i}\bra{i}_{\mathsf{B}}.
    \end{split}
    \label{eq:rhobbar}
\end{equation}
Calculating the entropy is straightforward, because the off-diagonal elements of $\rho_{\mathsf{B}}$ have averaged to zero, and, in the large $N$ approximation, gives

\begin{equation}
\boxed{
   S\left(\overline{\rho_{\mathsf{B}}}(t)\right) = -\overline{g}(t)\log \left(\overline{g}(t)\right) -\left(1-\overline{g}(t)\right) \log \left(\frac{1-\overline{g}(t)}{N}\right),
   }
    \label{eq: large N entropy average bath}
\end{equation}
At $t=0$, the entropy is zero; $\rho_{\mathsf{B}}(0) = \ket{0}\bra{0}_{\mathsf{B}}$. However, as $\overline{g}(t)$ decays as a function of time, the von Neumann entropy keeps rising. 
Taking the long time average, for which $\overline{g} \sim 1/N$, the entropy of the averaged state limits to
\begin{equation}
   \lim_{t \to \infty}S(\overline{\rho_{\mathsf{B}}}(t)) =\log N.
    \label{eq: na\"ive entropy}
\end{equation}
Contrary to the entropy of subsystem ${\mathsf{A}}$, the von Neumann entropy of the averaged environment state does not return to zero at late times. This might seem contradictory, as we had taken time evolution to be unitary. The initial state of the combined system $\mathsf{AB}$ is pure, so the entropy of the system $\mathsf{A}$ and the environment $\mathsf{B}$ must be equal at all times. This is reminiscent of the classical information paradox, where the radiation left after the black hole has evaporated seems to be mixed, after unitarily evolving a pure state.
Therefore, \eqref{eq: large N entropy average bath} is our analogue of the Hawking curve, where the radiation entropy grows and remains large at late times, signalling a mixed state.
This apparent contradiction can be resolved by realising that we have not studied the average entropy $\overline{S_{\mathsf{A}}}(t)$ and $\overline{S_{\mathsf{B}}}(t)$, but rather the entropy of the averaged states. 

\subsubsection{GUE}
If we specify which ensemble we are drawing our random Hamiltonians from, we can make more precise statements about the time evolution of the von Neumann entropy. 
We average over random Hamiltonians with the GUE probability measure, where we have taken \eqref{eq: Gaussian ensemble measure} with $a = \sqrt{2/N}$:
\begin{equation}
    \overline{f(H)} := \int dH \exp\left(-\frac{N}{2}\Tr H^2\right) f(H).
    \label{eq: measure}
\end{equation}
The choice of the prefactor $a$ in the exponent controls the width of the eigenvalue spectrum. Our choice gives an $N$-independent width, $(\overbar{H^2} - \overbar{H}^2) \sim 1$, which is what we want for our microcanonical energy windows of fixed width\footnote{We could include an extra dimensionless parameter $\alpha$ by replacing $N$ by $N\alpha^2$ in \ref{eq: measure}. This would be the same as rescaling time via $t\rightarrow t/\alpha$, and in this sense the unit of time can be chosen arbitrarily in our toy model.}. 
Other Gaussian ensembles are the GOE and the GSE, where the integral is over the space of real symmetric matrices and symmetric quaternionic matrices respectively, and the pre-factor of the measure is slightly altered.
We choose GUE, rather than GOE or GSE, because it does not assume time reversal symmetry and is the least restrictive symmetry class of the three Gaussian ensembles. We expect that all our results could be generalised to GOE or GSE if one wished to assume time reversal and/or rotational symmetry, and we could also consider quartic and higher order terms in the matrix potential.

\begin{figure}
    \centering
    \includegraphics[width=0.7\textwidth]{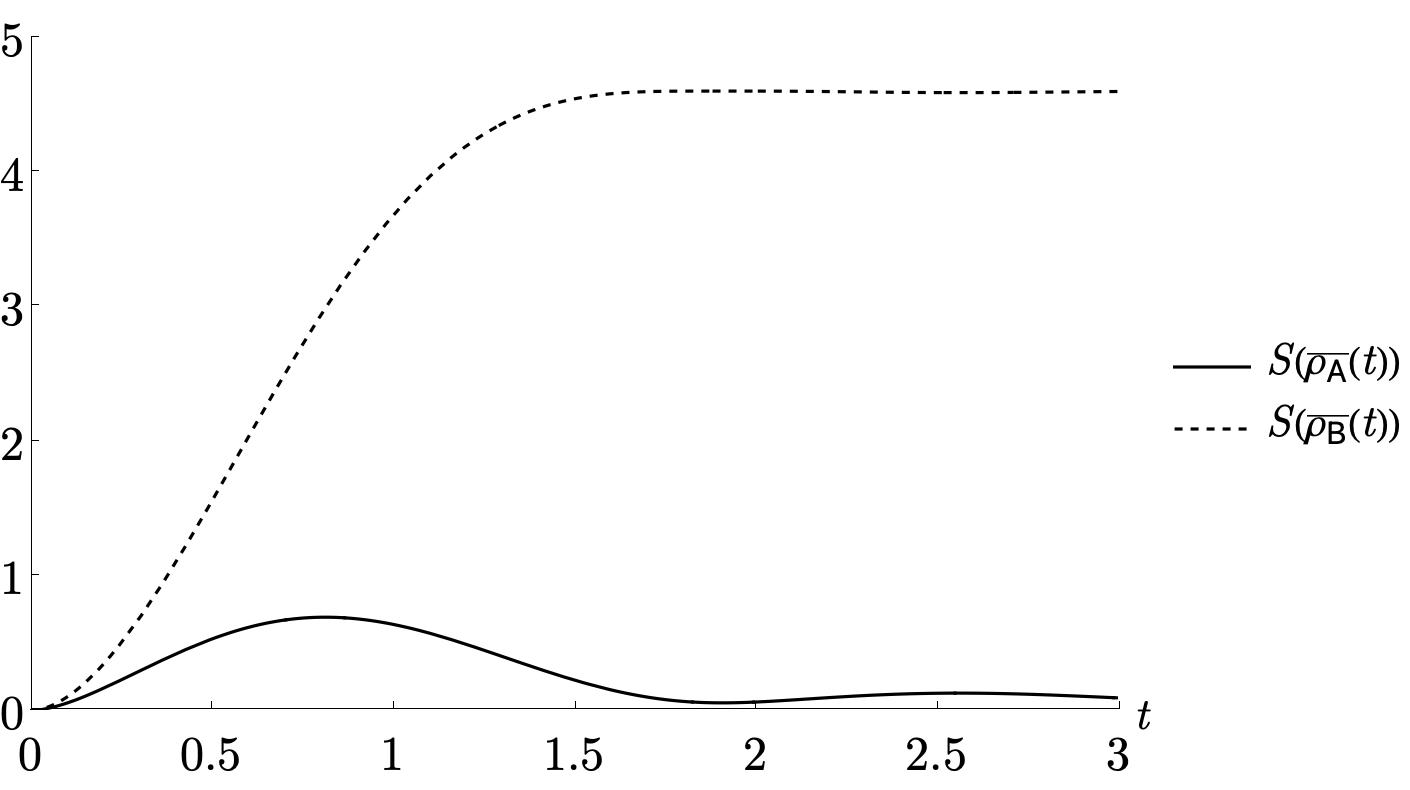}
    \caption{Entropy of the averaged system and environment density matrices. 
    The solid line shows $S(\overline{\rho_{\mathsf{A}}}(t))$, while the dashed line shows $S(\overline{\rho_{\mathsf{B}}}(t))$. 
    These are numerical plots that show the average over 10 Hamiltonians from the GUE with $N=100$. We have used the analytic results for the Haar averaged density matrices, \eqref{eq:rhoAbar} and \eqref{eq:rhobbar}, and plotted using the numerical average of the spectral form factor, which also agrees with the analytically averaged result.
   The entropy for the subsystem ${\mathsf{A}}$ grows to a maximum value of $\log 2$ and then decreases until it is approximately zero. The entropy for the subsystem ${\mathsf{B}}$ rises monotonically and approaches $\log N$ at late times. }
    \label{fig:naive entropy}
\end{figure}

For the GUE, \cite{Cotler2017chaos} studied the time evolution for the average spectral form factor, also using $a = \sqrt{2/N}$ as in \eqref{eq: measure}. They find the approximate\footnote{The approximation made to find this result is a short distance cutoff to regulate the divergence for closeby eigenvalues in the sine kernel. The divergence is an artefact of the expansion around infinite $N$. However, when comparing with numerics, \cite{Cotler2017chaos} finds that the difference between this approximate function and the numerical solution does not decrease when $N$ increases, meaning the approximation is not in effect a large $N$ approximation. This discrepancy only appears in the ramp; the slope and plateau are well approximated by \eqref{eq: W}.
} result
\begin{equation}
    g(t)N^2 = N^2 r_1(t)^2 -N r_2(t) + N,
    \label{eq: W}
\end{equation}
where the functions $r_1(t)$ and $r_2(t)$ are defined as
\begin{equation}
    r_1(t) \coloneqq  \frac{J_1(2t)}{t}, \qquad r_2(t) \coloneqq \begin{cases}
        1 - \frac{t}{2N}, \quad \text{for } t < 2N,\\
        0 \qquad \qquad \text{for } t > 2N.
    \end{cases}
\end{equation}
Here $J_1$ is a Bessel function of the first kind.
For Hamiltonians of the GUE with $N=100$, the time evolution of the von Neumann entropy of the averaged states is visualised in Fig. \ref{fig:naive entropy}. It can be seen that indeed the entropy of the subsystem $\mathsf{A}$ increases until it reaches its maximum $\log 2$, and later drops and becomes close to zero again. On the other hand, the entropy of the subsystem $\mathsf{B}$ keeps increasing, approaching $\log N$.

\subsection{Purity}
\label{sec: Average purity}
\subsubsection{Average purity of system and environment}
The fact that we have found a different von Neumann entropy for the averaged state in subsystem ${\mathsf{A}}$ than subsystem ${\mathsf{B}}$ implies the total averaged state in the full system $\mathsf{AB}$ is no longer pure. We explicitly compute this by averaging \eqref{eq: rho(t)} over the unitary matrices:
\begin{equation}
    \begin{split}
        \int_{U_N}\rho(t)
        &= \frac{1+g(t)N}{N+1}\ket{\psi_1}\bra{\psi_1} + \frac{N-g(t)N}{N^2-1} \sum_{i=2}^N\ket{\psi_i}\bra{\psi_i}.
    \end{split}
\end{equation}
If we take the square of this density matrix and then take the trace, we find that the  purity of the averaged state in the large $N$ approximation is given by
\begin{equation}
    \begin{split}
        \Tr \left(\overline{\rho}^2(t)\right)
        &= \overline{g}(t)^2 + \frac{
    1}{N}\left(1-\overline{g}(t)\right)^2.
    \end{split}
\end{equation}
Given the time evolution of the spectral form factor, we see that the purity of the averaged total state starts at 1, but then decreases and goes to $1/N$ in the large $N$ approximation. However, under unitary time evolution, the density matrix remains pure at all times. Indeed, computing the expression for the purity without averaging the state gives
\begin{equation}
\begin{split}
     \Tr (\rho^2 (t))
     &\equiv 1
\end{split}
\end{equation}
at all times. This follows from the properties of the unitary matrices and no averaging has come into play yet.

It is not immediately obvious what unitarity means when evolving with an ensemble of Hamiltonians. While evolution with a single Hamiltonian is unitary, the ensemble-averaged evolution of a given quantity may not be unitary in the sense of whether there exists a single unitary transformation that can give the same result as the ensemble average. The purity of $\overline{\rho}(t)$ that we calculated is not unitary in this sense.

\subsubsection{Average purity of system}
Let us now compute the averaged purity of the states reduced to $\mathsf{A}$ and $\mathsf{B}$. 
First, to compare, the purities of the averaged density matrices, in the large $N$ approximation, are
\begin{equation}
    \Tr \left(\overline{\rho_{\mathsf{A}}}^2(t)\right) = \overline{g}(t)^2+ \left(1 - \overline{g}(t)\right)^2, \qquad \Tr \left(\overline{\rho_{\mathsf{B}}}^2(t)\right) = \overline{g}(t)^2 + \frac{1}{N}\left(1-\overline{g}(t)\right)^2.
    \label{eq: purity average state}
\end{equation}
Therefore, at late times, the average of $\overline{\rho_{\mathsf{A}}}$ is pure, while $\overline{\rho_{\mathsf{B}}}$ has purity $\sim N^{-1}$. In contrast, at late times, the averaged purity of $\mathsf{B}$ should be approximately one. Assuming ergodicity, the long time average of the time-evolved state equals the microcanonical average and, since a large fraction $\frac{N-1}{N}$ of the states in $\cH_{\text{micro};E\pm \delta E}$ are in $\ket{0}_{\mathsf{A}} \otimes \cH_{\mathsf{B};E}$, the microcanonical average is approximately an unentangled product state with pure reduced density matrices. 
Therefore, we expect the average purity of both the system and environment at late times to be approximately one.
This differs from the purity of the averaged state in $\mathsf{B}$ as seen in \eqref{eq: purity average state}, which goes to zero at late times. We should then expect that, at late times, 
\bne \Tr \left(\overline{\rho_{\mathsf{B}}}^2\right) \neq \overline{\Tr \left(\rho_{\mathsf{B}}^2\right) }.\ene

As we will show next, the difference between the averaged purity and the purity of the averaged state comes from, when averaging over the unitary group, whether the unitary matrix index contractions connect the reduced density matrices or not. 
\subsubsection{Average purity of the environment}
Let us now study the average purity of the state reduced $\mathsf{B}$. 
Squaring the partial density matrix and taking the trace, we obtain
\begin{equation}
\begin{split}
    \Tr \left(\rho_{\mathsf{B}}^2(t) \right)
      &= \sum_{k,l,p,q = 1}^N e^{i(\lambda_{k} -\lambda_{l}+\lambda_{p}-\lambda_{q})t} \bigg(U_{1k}U_{k1}^\dagger U_{1l}U_{l1}^\dagger ~ U_{1p}U_{p 1}^\dagger U_{1q}U_{q1}^\dagger\\
     &\phantom{=}+\sum_{i,j = 2}^N 
      U_{ik}U_{k1}^\dagger U_{1l}U_{lj}^\dagger ~  U_{jp}U_{p 1}^\dagger U_{1q} U_{qi}^\dagger \bigg).
\end{split}
\end{equation}
Note that both of these terms are very similar, where $i=j=1$ for the first term, and for the second term both $i$ and $j$ can be anything but 1.
For both terms, the first four unitaries come from the left copy of the density matrix, the last four unitary matrices come from the right one. 
Taking the trace and averaging over the unitary matrices, there are a priori many different contraction patterns that need to be taken into account.\footnote{Averaging over the unitary matrices and taking the trace commute. The averaging is an integration and the trace is a sum which, because it is over a finite number of terms, commutes with the integration.}

\subsubsection*{Large $N$ approximation}
\begin{figure}
    \centering
    \includegraphics[width=\textwidth]{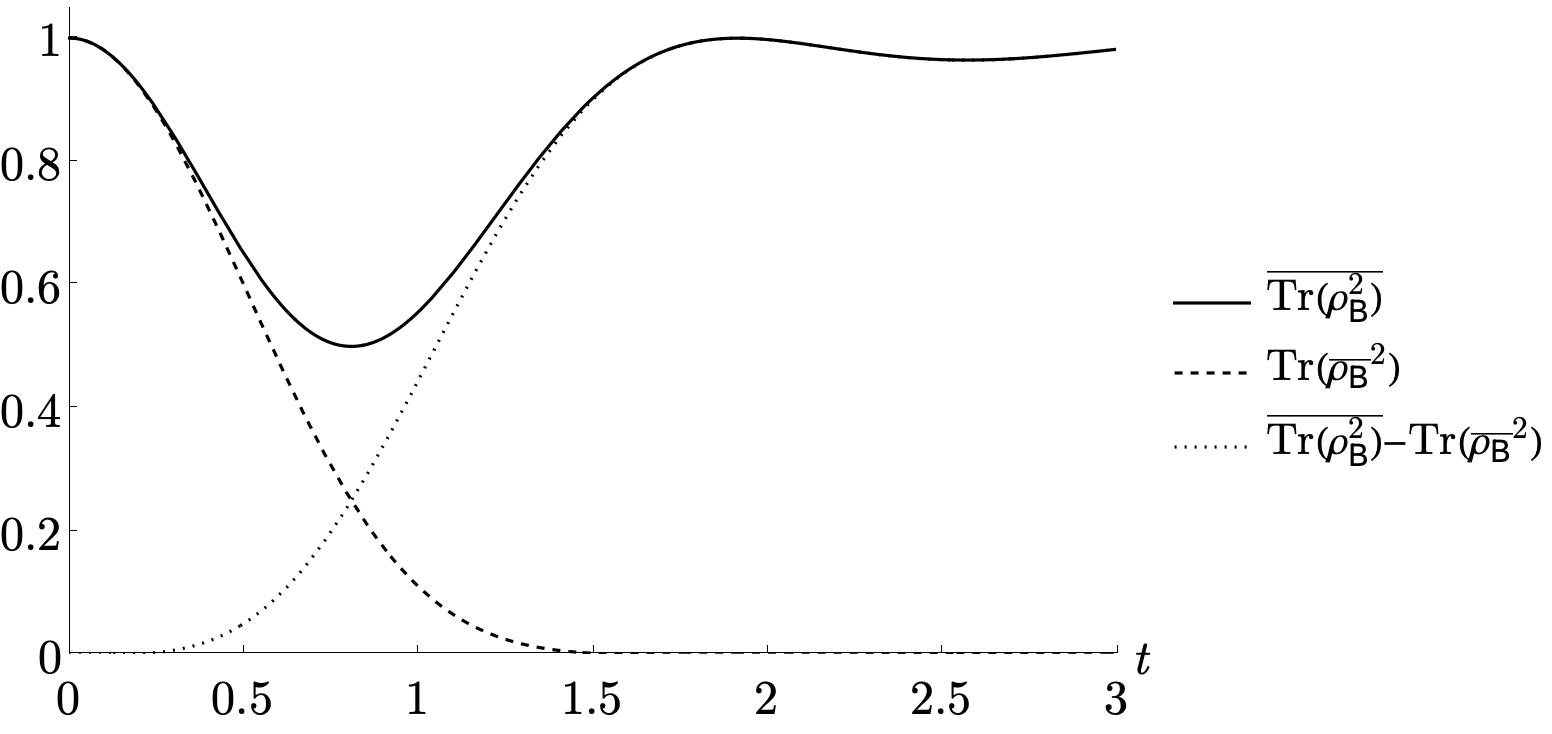}
    \caption{The purity
    of the environment $\mathsf{B}$ as a function of time. The solid black line is a numerical plot of the large $N$ approximation of $\overline{\Tr (\rho_{\mathsf{B}}^2(t))}$ as defined in \eqref{eq:purity} averaged over 10 Hamiltonians drawn from the GUE with $N =100$. 
    The dashed line is the disconnected component of $\overline{\Tr (\rho_{\mathsf{B}}^2(t))}$ and the dotted line is the connected component, as defined in \eqref{eq: disc and conn}. 
    At early times, the disconnected component dominates, while at late times the connected component dominates, resulting in a final state that is close to pure. This is closely analogous to the replica wormhole story, where unitarity in the Page curve is restored by an exchange of dominance to a saddle that connects the density matrices when calculating the R\'enyi entropies. 
    }
    \label{fig:purity GUE}
\end{figure}

Taking the Haar average of a product of unitary matrices contracts indices. To explain our contraction notation, we will draw the line above for contractions between unitary matrices and below for contractions between individual indices, so
\begin{equation}
\contraction[6pt]{}{U}{{}_{ab}}{U}
\bcontraction[6pt]{U_{ab}U_{cd}^\dagger \coloneqq U_{a}}{{}_{b\!\!}}{U}{{}_c}
\bcontraction[10pt]{U_{ab}U_{cd}^\dagger \coloneqq U}{{}_{a\!}}{U_c}{{}_{d\,}}
U_{ab}U_{cd}^\dagger \coloneqq U_{ab}U_{cd}^\dagger \sim \frac{\delta_{ad}\delta_{bc}}{N}.
\end{equation}
We will first work in the large $N$ approximation, where only five contraction patterns of the indices of the unitary matrices turn out to be non-vanishing. The first contraction pattern that needs to be taken into account is 
\begin{equation}
 \contraction[6 pt]{}{U}{{}_{ik}}{U} 
 \contraction[6pt]{U_{ik} U_{k1}^\dagger }{
 U}{{}_{1l}}{U}
\contraction[6pt]{U_{ik}U_{k1}^\dagger U_{1l}U_{lj}^\dagger \quad}{  U}{{}_{jp} }{U}
\contraction[6pt] {U_{ik}U_{k1}^\dagger U_{1l}U_{lj}^\dagger \quad U_{jp} U_{p 1}^\dagger}{  U}{{}_{1q}}{ U}
 U_{ik}U_{k1}^\dagger U_{1l}U_{lj}^\dagger \quad  U_{jp} U_{p 1}^\dagger  U_{1q} U_{qi}^\dagger \sim \frac{\delta_{i1}\delta_{j1}}{N^4} \longrightarrow \overline{g^2}(t).
 \label{eq: contraction 1}
\end{equation}
There is a space left between the unitary matrices that belong to the first and second copy of $\rho_{\mathsf{B}}$, to highlight whether the contraction connects the different copies of $\rho_{\mathsf{B}}$ or not. 
The delta functions give a contribution of $\overline{g^2}$ after evaluating the sums together with the exponential factor. Since $g$ is of order one at early times, we indeed have to include this term in our leading order expansion for the purity. 
There are four more contractions that contribute to leading order to the purity:
\begin{equation}
\begin{split}
\contraction[12 pt]{}{U}{{}_{ik} U_{k1}^\dagger U_{1l}U_{lj}^\dagger  \quad U_{jp} U_{p 1}^\dagger  U_{1q}}{U}
 \contraction[9pt]{U_{ik}}{U}{{}_{k1}^\dagger U_{1l}U_{lj}^\dagger  \quad U_{jp} U_{p 1}^\dagger}  {U} 
\contraction[6pt] {U_{ik}U_{k1}^\dagger}{ U}{{}_{1l}U_{lj}^\dagger \quad  U_{jp} }{U}
\contraction[3pt]{U_{ik}U_{k1}^\dagger U_{1l}}{U}{{}_{lj}^\dagger \quad }{ U}
 &U_{ik}U_{k1}^\dagger U_{1l}U_{lj}^\dagger \quad U_{jp} U_{p 1}^\dagger  U_{1q} U_{qi}^\dagger \sim \frac{\delta_{kq}\delta_{l p}}{N^4}  \longrightarrow 1,\\
 &\\
  \contraction[9pt]{U_{ik}}{U}{{}_{k1}^\dagger U_{1l}U_{lj}^\dagger \quad U_{jp} U_{p 1}^\dagger}  {U} 
 \contraction[12 pt]{}{U}{{}_{ik} U_{k1}^\dagger U_{1l}U_{lj}^\dagger  \quad U_{jp} U_{p 1}^\dagger  U_{1q}}{U}
\bcontraction[6pt] {U_{ik}U_{k1}^\dagger U_{1}}{{}_{l\!\!}}{U}{{}_{l}^\dagger}
  \bcontraction[6pt]{U_{ik}U_{k1}^\dagger U_{1l}U_{lj}^\dagger \quad  U_{j}}{{}_{p\!\!}}{U} {{}_{p \!\!\!\!\!\!\!\!\!\!\!}^\dagger}
  \bcontraction[6pt]{U_{ik}U_{k1}^\dagger U_{1l}U_{j\!\!\!\!\!}^\dagger}{{}_{j}\!\!\!}{  \quad U}{{}_{j}}
  \bcontraction[10pt]  { U_{ik}U_{k1}^\dagger U}{{}_{1}\!\!}{{}_{l}U_{l j}^\dagger \quad U_{jp} U_{p}}{{}_1\,}
     &U_{ik}U_{k1}^\dagger U_{1l}U_{l j}^\dagger \quad U_{jp} U_{p 1}^\dagger  U_{1q} U_{q1}^\dagger \sim \frac{-\delta_{kq}}{N^5}\longrightarrow -\overline{g}(t),\\
     &\\
\contraction[9pt]{U_{ik}U_{k1}^\dagger }{U}{{}_{1l} U_{l j}^\dagger \quad U_{jp}}{U}
\contraction[6pt]{U_{ik}U_{k1}^\dagger U_{1l}}{U}{{}_{lj}^\dagger \quad }{U}
\bcontraction[6pt]{U_{i}}{{}_{k}}{U^\dagger}{{}_{k\!\!\!\!\!\!\!\!\!\!}}
\bcontraction[6pt]{U_{ik}U_{k1}^\dagger U_{1l}U_{l j}^\dagger \quad U_{jp} U_{p 1}^\dagger  U_{1}}{{}_{q}} {U}{{}_{q \!\!\!}}
\bcontraction[6pt]{U_{ik}U_{k}^\dagger}{{}_1\!} {U_{1l}U_{l j}^\dagger  \quad U_{jp} U_{p 1}^\dagger  U}{{}_1} 
 \bcontraction[9pt]{U}{{}_{i}}{{}_{k\!\!}U_{k1}^\dagger U_{1l}U_{l j}^\dagger \quad U_{jp} U_{p 1}^\dagger  U_{1q} U_{q}}{{}_{i\!\!}}
    &U_{ik}U_{k 1}^\dagger U_{1l}U_{l j}^\dagger \quad U_{jp} U_{p 1}^\dagger  U_{1q} U_{qi}^\dagger \sim \frac{-\delta_{lp}}{N^5} \longrightarrow -\overline{g}(t),\\
    &\\
\bcontraction[5pt]{ U_{ik}U_{k 1}^\dagger U_{1}}{{}_{l}\!\!}{U^\dagger}{{}_{l}\!\!\!\!}
\bcontraction[5pt]{U_{ik}U_{k1}^\dagger U_{1l}U_{l}^\dagger}{{}_{j\!\!\!\!\!}}{\quad U}{{}_{j\,\,\,}} 
\bcontraction[5pt]{  U_{ik}U_{k1}^\dagger U_{1l}U_{l j}^\dagger \quad U_{j}}{{}_{p}\!}{U} {{}_{p}\!\!}
\bcontraction[8pt] {U_{ik}U_{k1}^\dagger U}{{}_1\!\!}{{}_{l}U_{l j}^\dagger  \quad U_{jp} U_{p }}{{}_1}
\bcontraction[13pt]{U_{ik}U_{k}^\dagger}{{}_1\!} {U_{1l}U_{l j}^\dagger  \quad U_{jp} U_{p 1}^\dagger  U}{{}_1} 
\bcontraction[13pt]{U_{i}}{{}_{k}}{U^\dagger}{{}_{k\!\!\!\!\!\!\!\!}}
\bcontraction[13pt]{U_{ik}U_{k1}^\dagger U_{1l}U_{l j}^\dagger \quad U_{jp} U_{p 1}^\dagger  U_{1}}{{}_{q}} {U}{{}_{q \!\!}}
 \bcontraction[16pt]{U}{{}_{i}}{{}_{k\!\!}U_{k1}^\dagger U_{1l}U_{l j}^\dagger \quad U_{jp} U_{p 1}^\dagger  U_{1q} U_{q}}{{}_{i}}
     &U_{ik}U_{k1}^\dagger U_{1l}U_{l j}^\dagger \quad U_{jp} U_{p 1}^\dagger  U_{1q} U_{qi}^\dagger \sim \frac{1}{N^6}  \longrightarrow \overline{g^2}(t).
     \label{eq: contraction 2}
\end{split}
\end{equation}
Therefore we find that to leading order in the large $N$ approximation, the purity of the environment is given by 
\begin{equation}
     \overline{\Tr (\rho_{\mathsf{B}}^2(t))} = 1 - 2 \overline{g}(t)+2 \overline{g^2}(t)\, .
    \label{eq:purity}
\end{equation}
At $t=0$, the purity is equal to one because $g(0) = 1$. At late times, in chaotic systems and in the large $N$ approximation, $g$ goes to zero at late times and the purity will become one again. 
We can disentangle the contributions to the purity into a disconnected component and a connected component, and find
\begin{equation}
    \begin{split}
         \overline{\Tr (\rho_{\mathsf{B}}^2(t))}_{\text{disc.}} = \overline{g^2}(t)\, , \qquad
      \overline{\Tr (\rho_{\mathsf{B}}^2(t))}_{\text{conn.}} = \overline{\left(1-g(t)\right)^2}\, .\\
    \end{split}
    \label{eq: disc and conn}
\end{equation}
The disconnected component of the purity as computed in \eqref{eq: purity average state} starts out at 1 but decreases to zero at late times. Therefore, the dominant contribution to the purity changes. At early times, the disconnected component dominates. However, at some intermediate time, the connected component starts to dominate.
For a fixed ensemble, one can actually compute the time evolution of the purity. For the GUE, the purity and its decomposition into disconnected and connected components is shown in Fig. \ref{fig:purity GUE}. 

Our model captures the replica wormhole story as there is a dynamical transition in dominance: the connected contribution grows larger than the disconnected contribution. This connected contribution contracts the density matrix indices exactly how we would expect it to in the replica wormhole:
\begin{equation}
    \bcontraction[10pt]{\rho}{{}_{b_1\!\!}}{_{b_2}\rho_{b_2}}{{}_{b_1\!\!}}
    \bcontraction[7pt]{\rho_{b_1}}{{}_{b_2\!\!}}{\rho}{{}_{b_2}}
    \rho_{b_1 b_2} \rho_{b_2 b_1}.
\end{equation}

Let us calculate the time at which the averaged purity of $\rho_{\mathsf{B}}$ is minimal, which is an analogue of a black hole's Page time in our model. The turnaround is when $\overline{g} = \frac{1}{2}$, where the purity has dropped to $\frac{1}{2}$.\footnote{For this, we assume that $\overline{g^2} \approx \overline{g}^2$ for $t \ll \sqrt{N}$. 
Spectral form factors in chaotic theories are characterised by their dip, ramp and plateau, and they are self-averaging before their dip time~\cite{prange1997spectral}, but not after. For the GUE ensemble, $t \sim \sqrt{N}$ is the dip time. 
Numerical evidence to justify our assumption is shown in Fig. \ref{fig: error}.}
For the GUE, taking $\overline{g}(t) \approx e^{-t^2/2}$ which is valid at early times, it can be seen that the transition time is 
\begin{equation} \label{eq:PageTimePurity}
    t \approx \sqrt{\ln 2}.
\end{equation}
This time is independent of $N$.
In Fig. \ref{fig:purity GUE} it can be seen that this agrees with the numerical result since $\sqrt{\ln 2}\approx 0.83.$
To compare, the Page time of a 4d asymptotically flat Schwarzschild black hole of initial energy $E$ is 
\bne \label{eq:tPage} t_{\text{Page}} \sim G_N^2 E^3. \ene
Our model has no parameters that directly represent the Planck scale, or $E$, the total energy of the system and environment, so cannot capture the dependence of the black hole entropy or the Page time on these quantities. Nonetheless, it does capture the qualitative shapes of the unitary and non-unitary Page curves. 
That said, in Sec. \ref{sec: Extensions} we will explain how our model can be modified to get a more physical Page time like~\eqref{eq:tPage}.

\subsubsection*{Exact in $N$}
One can also compute an expression for the averaged purity that is exact in $N$. To do so, one needs the exact Weingarten functions for 4-cycles, which are listed in App. \ref{app:weingarten}. The result is
\begin{equation}
  \overline{\Tr (\rho_{\mathsf{B}}^2(t))} =  \frac{N^4\left(1-2\overline{g}+2\overline{g^2}\right)+ N^3\left(4-2\overline{g}+2\overline{g_{3}}+2\overline{g_{3}^*}\right)+N^2\left(5+4\overline{g}+2\overline{g_{2}}\right) +6 N}{N (N+1) (N+2) (N+3)}\,.
  \label{eq: purity all orders}
\end{equation}
Here $g_{2}(t)$ and $g_{3}(t)$ are new functions that are defined as
\begin{equation}
    \begin{split}
        g_{2}(t) \coloneqq \frac{\Tr \left(e^{2iHt}\right) \Tr \left(e^{-2iHt}\right)}{N^2}\, , \qquad
        g_{3}(t) \coloneqq \frac{\Tr \left(e^{2iHt}\right) \Tr \left(e^{-iHt}\right)^2}{N^3}\, .
    \end{split}
\end{equation}
Both of these are of order one at early times, which is also their maximal value. Note that $g_2(t)$ is equal to $g(2t)$.
It is straightforward to see that the large $N$ approximation of \eqref{eq: purity all orders} indeed reduces to \eqref{eq:purity}. 

\subsubsection{Average purity of the system}
Let us next calculate the purity of the system $\mathsf{A}$.
Squaring $\rho_{\mathsf{A}}$ and taking the trace yields
\begin{equation}
\begin{split}
      \Tr \rho_{\mathsf{A}}^2(t) 
       &= \sum_{k,l,p,q = 1}^N e^{i(\lambda_{k} -\lambda_{l}+\lambda_{p}-\lambda_{q})t} \bigg(
      U_{1k}U_{k1}^\dagger U_{1l}U_{l1}^\dagger ~ U_{1p}U_{p 1}^\dagger U_{1q}U_{q1}^\dagger \\
      &\phantom{=}+\sum_{i,j = 2}^N U_{ik}U_{k1}^\dagger U_{1l}U_{lj}^\dagger ~ U_{jp}U_{p 1}^\dagger U_{1q}U_{qi}^\dagger \bigg).
      \label{eq: purity A}
\end{split}
\end{equation}
The unitaries in~\eqref{eq: purity A} are exactly the same as the unitaries that appeared in the purity of the environment. 
Note that $\Tr({\overline{\rho_{\mathsf{A}}}^2(t)}) = \overline{\Tr{(\rho_{\mathsf{A}}^2(t)})}$, in the large $N$ approximation, because the system $A$ evolves towards its pure vacuum state for typical draws from the Hamiltonian ensemble, so its purity is self-averaging. In contrast to the environment $\mathsf{B}$, no replica wormholes are necessary for unitarity in the purity of $\mathsf{A}$. A small consequence of $\Tr({\overline{\rho_{\mathsf{A}}}^2(t)}) = \overline{\Tr{(\rho_{\mathsf{A}}^2(t)})}$ is that $\Tr({\overline{\rho_{\mathsf{A}}}^2(t)})= \overline{\Tr({\rho_{\mathsf{B}}^2}(t))}$ and we will return to this peculiarity in Sec. \ref{sec: Extensions}.

\subsection{Averaged R\'enyi and von Neumann entropies}
\label{sec: Renyi entropies and von Neumann entropy}

In the previous subsection, we calculated and discussed the evolution of the purities of the system ${\mathsf{A}}$ and its environment ${\mathsf{B}}$. 
Next, we would like to extend the analysis to von Neumann and the higher R\'enyi entropies. 
It would be interesting to see whether the results for the purity extend to higher $n$, in particular whether the analogue of the dominant replica wormhole in \cite{Penington2022replica} is also the late time dominant contribution in our model and if so, when this exchange in dominance takes place.

In $\Tr(\rho_{\mathsf{X}}^n)$, whether for $\rho_{\mathsf{X}} = \rho_{\mathsf{A}}, \rho_{\mathsf{B}}$ or $\rho$, each density matrix contributes four unitary matrices, see~\eqref{eq: rho(t)}-\eqref{eq: rhoB}, 
so the Haar average of $\Tr(\rho_{\mathsf{X}}^n)$ is an integral over $4n$ unitary matrices with in total $8n$ indices. 
After averaging, all contraction patterns that are non-vanishing in the large $N$ approximation have their $8n$ indices contracted in pairs.\footnote{By contracted in pairs we mean the following. Each summed over index $i$ appears twice. In the leading order contractions, each $i$ is either contracted to itself, or both $i$'s are contracted with the two occurrences of the same index $j$. }
These pairwise contractions either connect different density matrices or not. A key result in this subsection that we will show is how connected contributions to the Haar average of $\Tr\left({\rho_{\mathsf{B}}^n}\right)$ restores unitarity in the Page curve.

\subsubsection{Disconnected contributions}

We can calculate the averaged von Neumann entropy through\footnote{Strictly speaking, we should average after taking the derivative, rather than the other way around. We will assume that these are the same; this seems like a mild assumption because $\rho_{\mathsf{B}}$ is a finite size matrix with eigenvalues in the compact interval $[0,1]$, so we expect nice convergence and analyticity properties.  
} 
\begin{equation}
    \overline{S(\rho_{\mathsf{B}}(t))} 
    = - \partial_n \overline{\Tr \left({\rho_{\mathsf{B}}^n(t)}\right)}\bigg \rvert_{n=1}.
\end{equation}
The expression for $\Tr \left({\rho_{\mathsf{B}}^n}\right)$ can be compactly written as
\begin{equation}
    \begin{split}
    \Tr \left({\rho_{\mathsf{B}}^n}\right)&= 
    \sum_{k_1=1}^N \dots \sum_{k_{2n}=1}^N ~ 
    \sum_{b_1=1}^N \dots \sum_{b_n=1}^N 
    \left(\prod_{m=1}^n \delta_{1b_m} + \prod_{m=1}^n (1-\delta_{1b_m})\right)\times\\
  &\phantom{=}\times \prod_{i=1}^n  
   e^{i(\lambda_{k_{2i-1}}-\lambda_{k_{2i}})t} U_{b_i k_{2i-1}} U_{k_{2i-1} 1}^\dagger U_{1 k_{2i}}  U_{k_{2i} b_{i+1}}^\dagger,
    \label{eq: trace n}  
    \end{split}
\end{equation}
where $b_{n+1} = b_1$. Note that there are two distinct terms in \eqref{eq: trace n}: the first term has $b_\ell=1$ for all $\ell$, while the second term has $b_{\ell} \neq 1$ for all $\ell$.

\begin{figure}
\centering
     \begin{subfigure}[b]{0.3\textwidth}
         \centering
         \includegraphics[width=\textwidth]{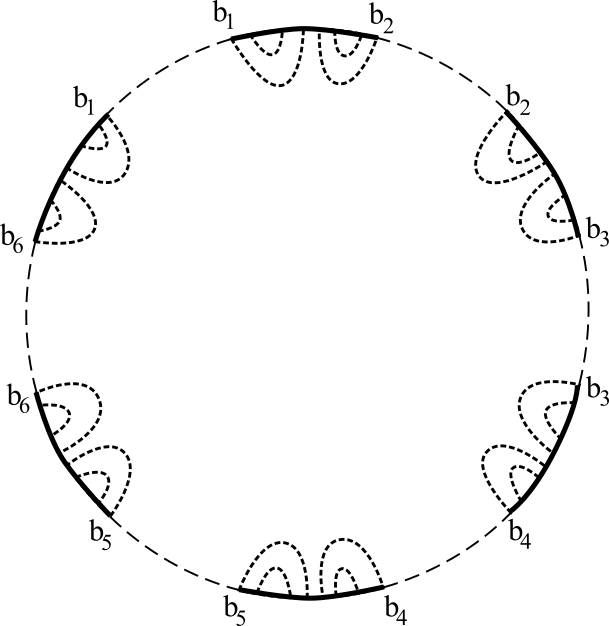}
    \caption{}
    \label{fig: disconnected1}
     \end{subfigure}
     \hfill 
     \begin{subfigure}[b]{0.3\textwidth}
         \centering
         \includegraphics[width=\textwidth]{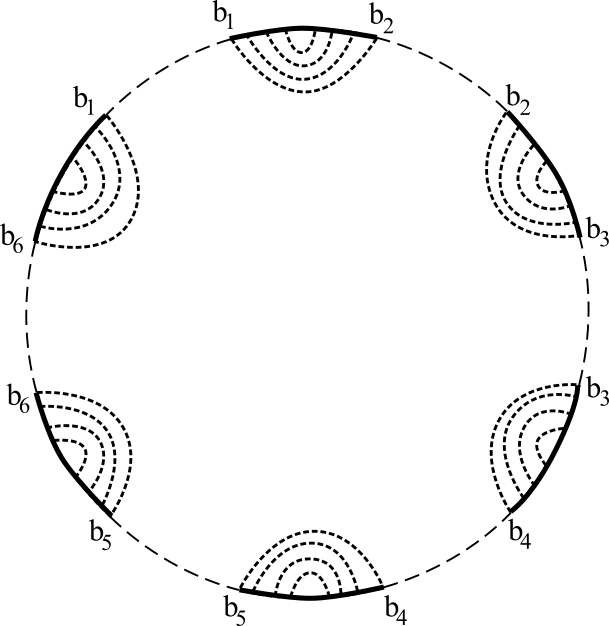}
         \caption{}
    \label{fig: disconnected2} 
     \end{subfigure}
     \hfill
     \begin{subfigure}[b]{0.3\textwidth}
         \centering
         \includegraphics[width=\textwidth]{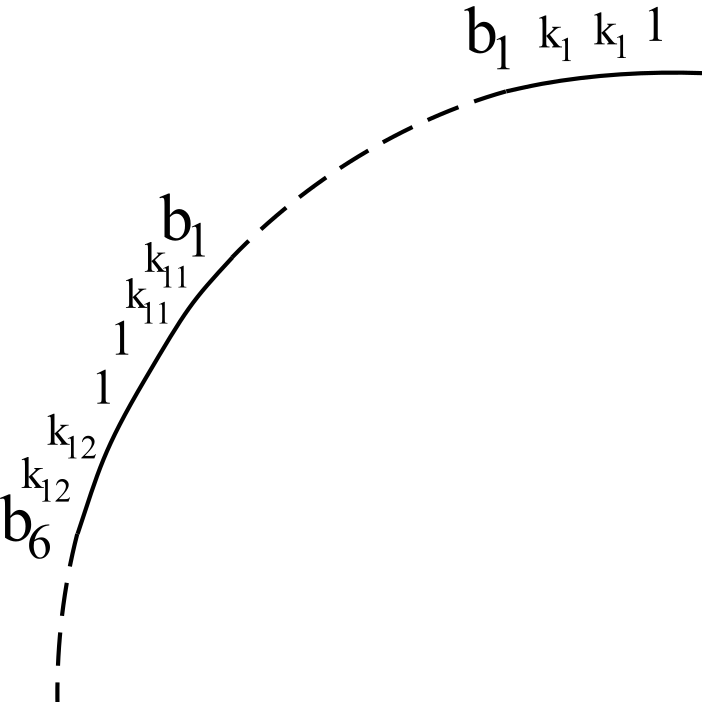}
    \caption{}
    \label{fig: disc zoomed}
     \end{subfigure}
\caption{Diagrammatics of disconnected contributions to the R\'enyi entropy, whose sum gives~\eqref{eq: large N disconnected}. The black lines on the edges of the first two circles represent the density matrices in $\Tr({\rho_{\mathsf{B}}^n(t)})$, and the density matrix indices are labelled. In \ref{fig: disconnected1} and \ref{fig: disconnected2}, two different contraction patterns of the unitary matrices contributing to \eqref{eq: large N disconnected} are visualised for $n=6$. The contraction pattern in \ref{fig: disconnected1} corresponds to the contribution in \eqref{eq: disc early time leading} that gives $\overline{g^6}(t)$. At early times, this is 1, but at late times it decreases to $\sim N^{-6}$. The contraction in \ref{fig: disconnected2} contributes $N^{-5}$. At early times, this is subleading, but at late times it dominates over \ref{fig: disconnected1}. In \ref{fig: disc zoomed}, a zoomed-in version on the upper left corner shows which lines correspond to which indices.
} 
\label{fig: disconnected wormholes}
\end{figure}

We will first study the disconnected contribution to $\overline{\Tr \left({\rho_{\mathsf{B}}^n}\right)}$, before we also take into account the connected components.
Each disconnected density matrix $\rho_{\mathsf{B}}$ has four possibilities to contract its indices, to leading order in the large $N$ approximation given by 
\begin{equation}
\overline{ U_{b_i k_{2i-1}} U_{k_{2i-1} 1}^\dagger U_{1k_{2i}} U_{k_{2i} b_{i+1}}^\dagger} =   \frac{\delta_{1b_{i}}\delta_{1 b_{i+1}}}{N^2}\left(1 - \frac{\delta_{k_{2i-1} k_{2i}}}{N}\right) + \frac{\delta_{b_i b_{i+1}}}{N^2}\left(\delta_{k_{2i-1} k_{2i}} - \frac{1}{N}\right).
\end{equation}
Therefore, we find that the leading order contribution of the disconnected component to $\overline{\Tr \left({\rho_{\mathsf{B}}^n}\right)}$ yields
\begin{equation}
    \overline{\Tr \left(\rho_{\mathsf{B}}^n\right)}_{\text{disc.}} = \overline{g^n}(t)+ \frac{\overline{\left(1-g(t)\right)^n}}{N^{n-1}} ~.
    \label{eq: large N disconnected}
\end{equation}
The first term in \eqref{eq: large N disconnected} corresponds to the contraction
\begin{equation}
    \prod_{i=1}^n \frac{\delta_{1 b_i}}{N^2}~.
    \label{eq: disc early time leading}
\end{equation}
At late times, $\overline{g} \to 1/N$. Therefore, the contribution of this term is of order $N^{-n}$ at late times. This contraction is visualised in Fig. \ref{fig: disconnected1} for $n=6$.
The second term in \eqref{eq: large N disconnected} corresponds to the contractions
\begin{equation}
    \prod_{i=1}^n \frac{\delta_{b_i b_{i+1}}}{N^2}\left(\delta_{k_{2i-1} k_{2i}} - \frac{1}{N}\right).
    \label{eq: disc late time leading}
\end{equation}
At late times, this is of order $N^{-n+1}$ and thus dominates over \eqref{eq: disc early time leading}. The contraction that survives the large $N$ approximation at late times and thus contributes $N^{-n+1}$ is visualised in Fig. \ref{fig: disconnected2} for $n=6$.

\begin{figure}
    \centering
    \begin{subfigure}[b]{0.45\textwidth}
         \centering
         \includegraphics[width=\textwidth]{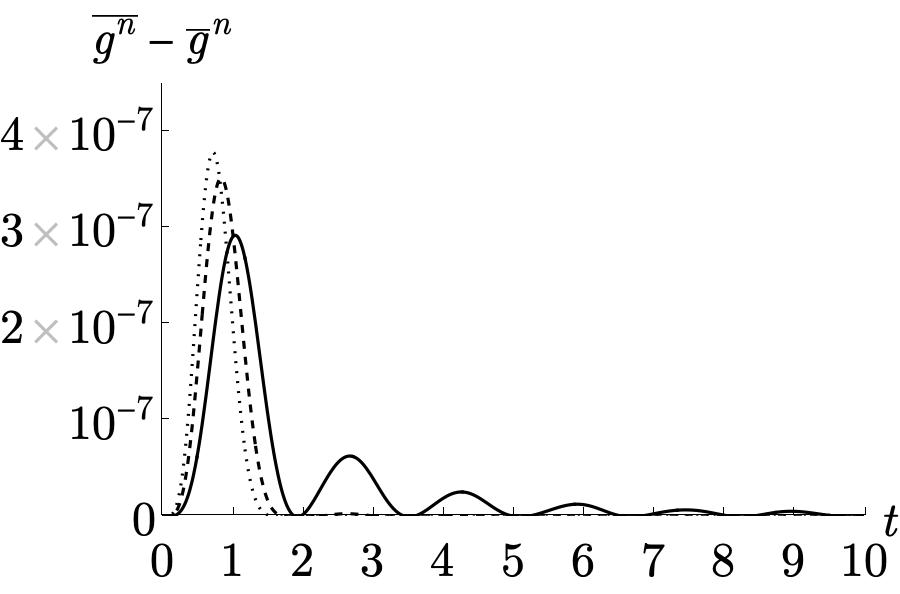}
        \caption{}
        \label{fig: abs}
     \end{subfigure}
     \hfill
    \begin{subfigure}[b]{0.45\textwidth}
         \centering
         \includegraphics[width=\textwidth]{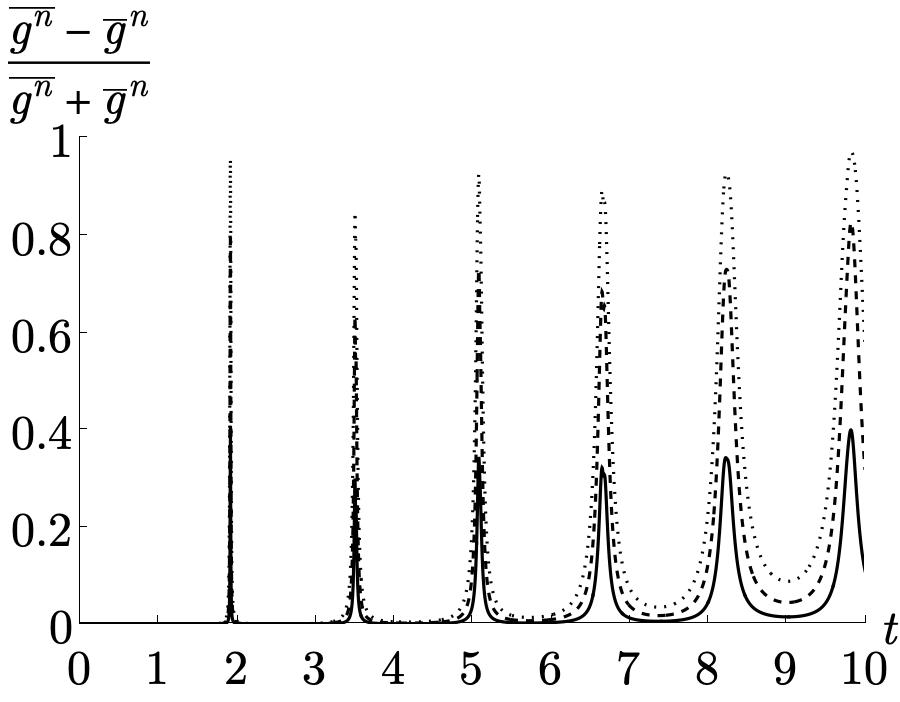}
        \caption{}
        \label{fig: rel long}
    \end{subfigure}
    \caption{In \ref{fig: abs}, the error $\overline{g^n}-\overline{g}^n$ is shown for $n=2,3,4$. The solid lines correspond to $n=2$, the dashed lines to $n=3$ and the dotted lines to $n=4$. The average is taken over 100 Hamiltonians drawn from the GUE with $N=1000$. In \ref{fig: rel long}, the relative error $\frac{\overline{g^n}-\overline{g}^n}{\overline{g^n}+\overline{g}^n}$ is visualised. It can be seen that the relative error has large spikes, centred around the zeroes of $J_1(2t)$ where $\overline{g}$ is expected to be very small before the dip time. We have numerical evidence that as $N$ increases, the width of these spikes decreases.}
    \label{fig: error}
\end{figure}

Spectral form factors are self-averaging before their dip times~\cite{prange1997spectral}. For GUE, the dip time is $t \sim \sqrt{N}$, which is long after the Page time of our model, so we can assume that $g$ is self-averaging and use the approximation \bne \label{eq: wn}\overline{g^n}(t) \approx \overline{g}^n(t) .\ene 
In Fig. \ref{fig: error} we give a numerical plot to show both the absolute and the relative error, to give further evidence for the approximation.

The leading order von Neumann entropy, only taking into account the disconnected components, can now be computed from \eqref{eq: large N disconnected} to be
\begin{equation}
\begin{split}
 \overline{S(\rho_{\mathsf{B}}(t))}_{\text{disc.}} 
      &\approx -\overline{g}(t)\log \left(\overline{g}(t)\right) -\left(1-\overline{g}(t)\right)\log \left(1-\overline{g}(t)\right) + \left(1-\overline{g}(t)\right)\log N\, .
\label{eq:discdisc}
\end{split}
\end{equation}
At late times, this approaches $\log N$. This is the same as we found for $S(\overline{\rho_{\mathsf{B}}}(t))$ in~\eqref{eq: large N entropy average bath} and thus at times when the assumption \eqref{eq: wn} is appropriate,
\bne
\overline{S(\rho_{\mathsf{B}})}_{\text{disc.}} \approx S(\overline{\rho_{\mathsf{B}}}).
\ene
The assumption in \eqref{eq: wn} is known to be valid up to the dip time, therefore the analytic continuation to the von Neumann entropy is only valid up until the dip time as well. At this point, we are ignorant about the behaviour after the dip time, both for the $n$'th R\'enyi entropies and the von Neumann entropy.

\subsubsection{Connected contributions}

\begin{figure}
\centering
     \begin{subfigure}[b]{0.3\textwidth}
         \centering
         \includegraphics[width=\textwidth]{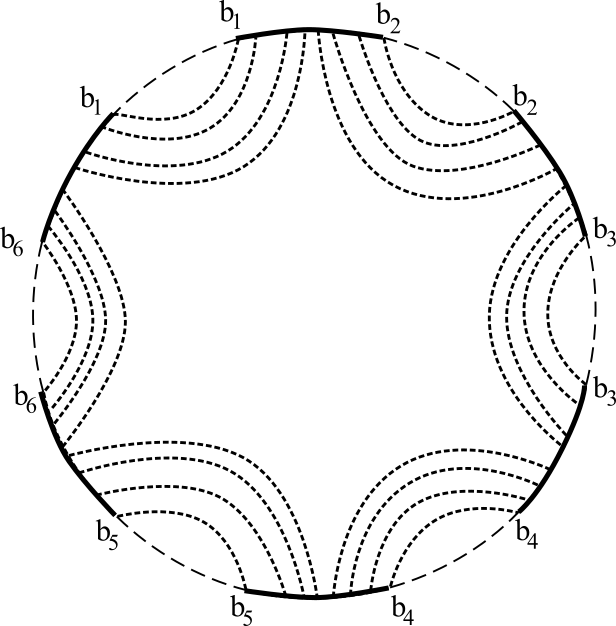}
    \caption{}
    \label{fig: fully con}
     \end{subfigure}
     \hfill
     \begin{subfigure}[b]{0.3\textwidth}
         \centering
         \includegraphics[width=\textwidth]{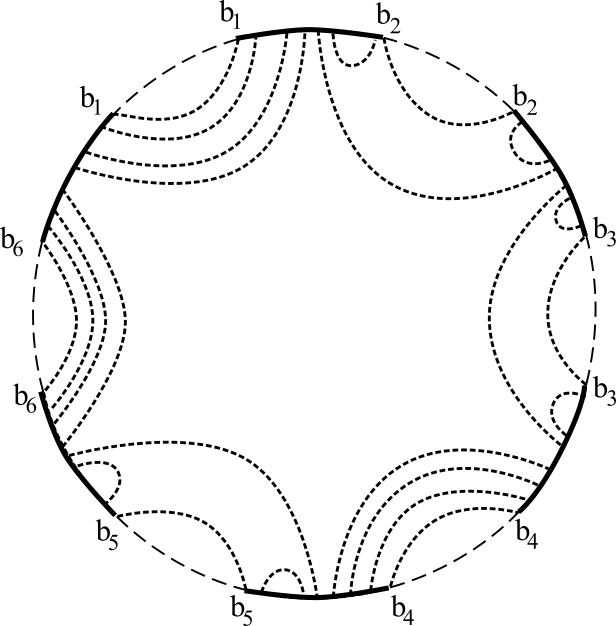}
    \caption{}
    \label{fig: partially con}
     \end{subfigure}
     \hfill 
     \begin{subfigure}[b]{0.3\textwidth}
         \centering
         \includegraphics[width=\textwidth]{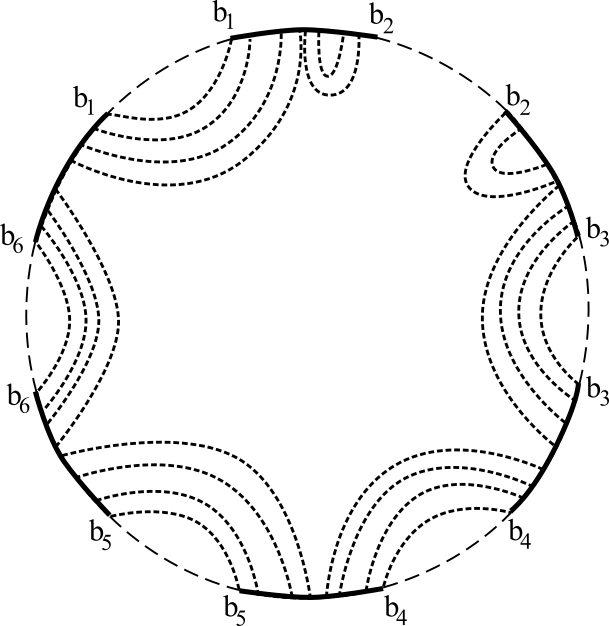}
         \caption{}
    \label{fig: subleading} 
     \end{subfigure}
\caption{Connected contributions to the R\'enyi entropy of the unitary matrix contractions. In \ref{fig: fully con}, \ref{fig: partially con} and \ref{fig: subleading}, three contraction patterns contributing to \eqref{eq: connected rho} are visualised. The contraction in \ref{fig: fully con} contributes 1, the contraction in \ref{fig: partially con} contributes $-\overline{g^3}(t)$. The contraction in \ref{fig: subleading} contributes $\frac{\overline{g}(t)}{N^5}$ and is always subleading with respect to the others.  At early times, \ref{fig: fully con} and \ref{fig: partially con} are of the same order, but at late times the only contraction pattern of the unitary matrices that is non-vanishing is the one in \ref{fig: fully con}.} 
\label{fig: unitary wormholes}
\end{figure}

\noindent Now we return to the computation of $\overline{\Tr \rho_{\mathsf{B}}^n}$, also taking into account the connected components. This yields
\begin{equation}
    \begin{split}
        \overline{\Tr \left(\rho_{\mathsf{B}}^n(t)\right)} 
        &= \left(1 -\overline{g}(t)\right)^n + \overline{g}^n(t).
        \label{eq: higher renyi}
    \end{split}
\end{equation}
Here we have again used the approximation \eqref{eq: wn}.
The computation of \eqref{eq: higher renyi} can be found in App. \ref{app: higher renyi}. 
The first term is the leading order contribution of the connected components, the second term corresponds to the disconnected component. We have dropped the terms corresponding to \eqref{eq: disc late time leading}, as they are always subleading with respect to the connected components for $n>1$. 
Three contributions to the connected components are visualised in Fig. \ref{fig: unitary wormholes}. 
We note that at $t=0$, the disconnected component is 1 and the sum of the connected components is zero. At late times, the disconnected component goes to zero, while the sum of the connected components approaches 1.
From \eqref{eq: higher renyi}, the large $N$ approximation of the $n$'th R\'enyi entropy is given by\footnote{
The averaged $n$'th R\'enyi entropy is proportional to $\overline{\log \Tr \left(\rho_{{\mathsf{B}}}^n\right)}$
but, as is well-known in the disorder-averaging literature, averaging after taking the logarithm is difficult so we instead average first and then take the logarithm. We thus are computing the annealed version instead of the quenched version of the R\'enyi entropies.
}
\begin{equation}
 \overline{S^{(n)}(\rho_{\mathsf{B}}(t))} = \frac{\log \left(\left(1 - \overline{g}(t)\right)^n + \overline{g}^n(t)\right)}{1-n}~, \qquad n > 1.
   \
\end{equation}
\subsection{Relation to replica wormholes}
\begin{figure}
\centering
     \begin{subfigure}[b]{0.4\textwidth}
         \centering
         \includegraphics[width=\textwidth]{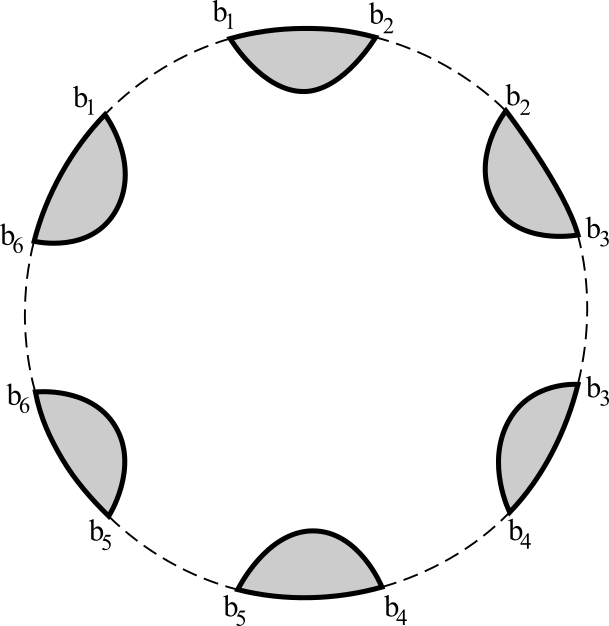}
    \caption{}
    \label{fig: disconnected rho}
     \end{subfigure}
     \hfill 
     \begin{subfigure}[b]{0.4\textwidth}
         \centering
         \includegraphics[width=\textwidth]{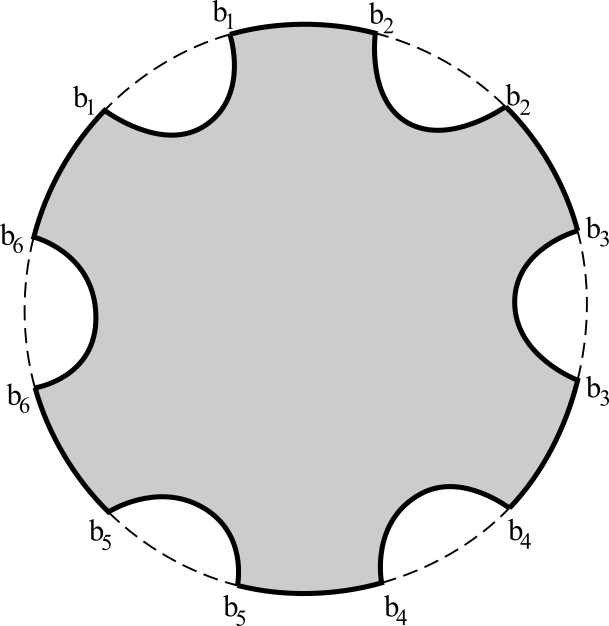}
         \caption{}
    \label{fig: connected rho} 
     \end{subfigure}
\caption{
The replica wormhole-like connectedness between density matrices that results from the unitary matrix contractions.
In \ref{fig: disconnected rho}, the disconnected contraction pattern from \eqref{eq: disconnected rho} is visualised. This pattern dominates at early times where it is close to 1, but approaches zero at late times. In \ref{fig: connected rho}, the contraction pattern corresponding to \eqref{eq: connected rho} is visualised. This term is close to zero at early times, but is the dominant contribution and approaches 1 at late times.} 
\label{fig: wormholes}
\end{figure}

In order to make contact with the replica wormholes as found in \cite{Penington2022replica}, we have to study the contractions of the density matrices $\rho_{\mathsf{B}}$. Each of these density matrices has two indices. We can thus write
\begin{equation}
    \Tr \left(\rho_{\mathsf{B}}^n\right) = \rho_{b_1 b_2} \rho_{b_2 b_3} \rho_{b_3 b_4} \dots \rho_{b_n b_1}.
    \label{eq: higher renyi 2}
\end{equation}
To avoid clutter, we have suppressed the ${\mathsf{B}}$-index of the partial density matrix.
Two examples of contributions to the disconnected components were given in Fig. \ref{fig: disconnected wormholes}. 
In terms of the contraction pattern of the density matrices, we can visualise the sum of these unitary contraction patterns as
\begin{equation}
    \bcontraction[6pt]{\rho}{{}_{b_1\!}}{}{b_2\!\!}
    \bcontraction[6pt]{\rho_{b_1 b_2}\rho}{{}_{b_2}\!\!}{}{b_3\!}
        \bcontraction[6pt]{\rho_{b_1b_2}\rho_{b_2b_3}\rho}{{}_{b_3\!\!}}{}{b_4\!\!}
    \bcontraction[6pt]{\rho_{b_1b_2} \rho_{b_2  b_3} \rho_{b_3b_4}\dots \rho}{{}_{b_n\!\!}}{}{b_1\!\!}
    \rho_{b_1b_2}\rho_{b_2b_3}\rho_{b_3 b_4}\dots \rho_{b_n b_1} = \overline{g^n}(t)+\frac{\overline{\left(1-g(t)\right)^n}}{N^{n-1}},
    \label{eq: disconnected rho}
\end{equation}
where each of the density matrices is contracted with itself. This contribution is 1 at $t=0$, but decreases as time evolves. This contraction of the density matrices is visualised in Fig. \ref{fig: disconnected rho} for $n=6$. 

On the other hand, in all of the leading order connected unitary contraction patterns, each $b_i$ is contracted with itself. Two examples of these contraction patterns are visualised in Fig. \ref{fig: fully con} and Fig. \ref{fig: partially con}.
Although the contraction of the internal indices of the unitary matrices varies between these different contributions, on the level of the density matrices the sum of the leading order contractions to the connected components can be visualised as 
\begin{equation}
    \bcontraction[9pt]{\rho}{{}_{b_1}\!}{{}_{b_2} \rho_{b_2 b_3} \rho_{b_3b_4} \dots \rho_{b_n}}{{}_{b_1\!\!}}
    \bcontraction[6pt]{\rho_{b_1}}{{}_{b_2}\!\!}{\rho}{{}_{b_2\!}}
    \bcontraction[6pt]{\rho_{b_1b_2}\rho_{b_2}}{{}_{b_3\!\!}}{\rho}{{}_{b_3\!\!}}
    \bcontraction[6pt]{\rho_{b_1b_2}\rho_{b_2b_3}\rho_{b_3}}{{}_{b_4\!\!}}{}{\dots\!\!\!}
    \bcontraction[6pt]{\rho_{b_1b_2}\rho_{b_2b_3}\rho_{b_3b_4}}{\dots\,\,\,\,\,\,}{\rho}{{}_{b_n\!\!\!\!\!\!\!\!\!\!\!\!\!\!\!\!\!\!\!\!}}
    \rho_{b_1b_2}\rho_{b_2b_3}\rho_{b_3b_4}\dots \rho_{b_n b_1} = \overline{\left(1-g(t)\right)^n}.
    \label{eq: connected rho}
\end{equation}
and pictorially as in Fig. \ref{fig: connected rho}.
This contribution is zero at $t=0$, but increases as time evolves until it approaches 1 at late times.
Note that there are other contraction patterns of the unitary matrices which may be harder to visualise as contraction patterns of the partial density matrices. An example of such a unitary contraction is given in Fig. \ref{fig: subleading}. However, all the leading order contractions can all be visualised as in Fig. \ref{fig: connected rho}. This is reminiscent of the leading order replica wormhole contribution as found in \cite{Penington2022replica}.

Now one can compute the large $N$ approximation of the von Neumann entropy of the environment $\mathsf{B}$ from \eqref{eq: higher renyi} to be\footnote{To remind the reader, in deriving this equation we assume $\overline{g^n}\approx \overline{g}^n$. Furthermore, we average before taking the derivative of $\Tr(\rho_{\mathsf{B}}^n)$ instead of after.}
\begin{equation}
\boxed{
    \overline{S(\rho_{\mathsf{B}}(t))}
     = -\overline{g}(t)\log \left(\overline{g}(t)\right) - \left(1-\overline{g}(t)\right)\log\left(1-\overline{g}(t)\right).
     \label{eq: von Neumann}
}
\end{equation}

\begin{figure}
    \centering
\includegraphics[width=0.7\textwidth]{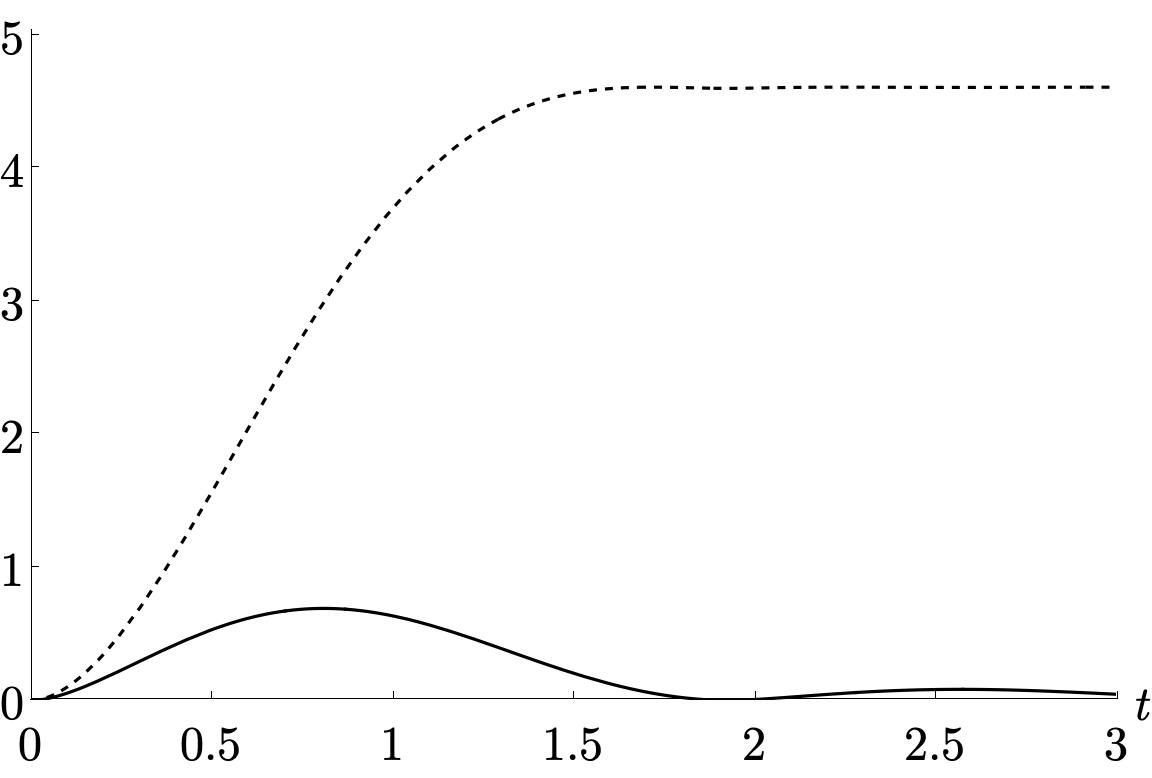}
    \caption{Time evolution of the large $N$ limit of the von Neumann entropy of the environment \eqref{eq: von Neumann}, averaged over 10 Hamiltonians drawn from the GUE with $N=100$. The dashed line is the von Neumann entropy of the disconnected components \eqref{eq:discdisc} and
    the solid line is the average von Neumann entropy \eqref{eq: von Neumann}.}
    \label{fig: von Neumann}
\end{figure}
The time evolution of the large $N$ von Neumann entropy is visualised in Fig. \ref{fig: von Neumann}. This is the unitary Page curve of our model.

It might be somewhat surprising to see that we have been able to drop $1/N$ subleading terms in the R\'enyi entropies and were able to retrieve the unitary result for the von Neumann entropy. It is not obvious that the large $N$ approximation and the replica limit of the R\'enyi entropies commute, i.e. that
\begin{equation}
   \lim_{N \to \infty} \lim_{n\to 1} \overline{S^{(n)}(\rho_{\mathsf{B}}(t))} =   \lim_{n\to 1} \lim_{N \to \infty}\overline{S^{(n)}(\rho_{\mathsf{B}}(t))}. 
\end{equation}
It is clear that the LHS should give the unitary result, as that is the large $N$ limit of the analytic continuation of the exact result. However, it is not obviously trivial that the RHS gives the same result; we are taking the leading order in $1/N$ for $n>1$ and taking the analytic continuation. In a semiclassical computation of the replica geometries, one also neglects the higher order corrections, but the analytic continuation of this approximate result still gives the correct von Neumann entropy. The `unreasonable effectiveness of the low energy EFT' in semiclassical gravity thus manifests itself through this mechanism in our model. 

In our model, the multi-boundary replica wormhole shown in Fig.~\ref{fig: connected rho}
is a representation of the class of unitary matrix index contractions that connect the density matrices in the way shown in Fig.~\ref{fig: fully con} and ~\ref{fig: partially con}. The connectedness arises from Haar averaging the unitary matrices coming from the random Hamiltonians. To summarise with a tongue-in-cheek slogan: ER = Haar.

\section{Extensions and further applications of the model}
\label{sec: Extensions}
In the previous section, we have seen how in a rather simple model we have captured several important qualitative features apparent in the black hole paradox, such as the na\"ive Hawking curve and the unitary Page curve, how replica wormholes are necessary to restore unitarity at late times and how the leading order contribution at late times is the cyclically connected replica wormhole. In this section, we will discuss some surprising features of our model, some improvements that can be made to make the model more realistic and we will look at the mutual information between the system, the environment and some reference system.

\subsection*{Difference with actual Page plot}
In the standard Page computation, the Hawking and Page curves follow each other at early times and only start to deviate around the Page time, when the Hawking curve keeps growing and the Page curve drops back down. Our result is different in this aspect: we have
\begin{equation}
\overline{S(\rho_{\mathsf{B}}(t))} - S(\overline{\rho_{\mathsf{B}}}(t)) = \left(1-\overline{g}(t)\right)\log N.
\end{equation}
This means the two curves start deviating immediately, and around the Page time already have a difference of $\frac{1}{2}\log N$. 
A curve which follows this feature of the Page curve a bit more closely is the quantity
\begin{equation}
    1 - \Tr (\rho_{\mathsf{B}}^2(t)).
    \label{eq: impurity}
\end{equation}
The difference between the disconnected and connected component of this ``impurity'' is visualised in Fig. \ref{fig: Page analogue}.

\subsection*{No need for replicas in the system}
 Another perhaps surprising feature of our model is that in the large $N$ approximation, 
\begin{equation}
    \overline{S(\rho_{\mathsf{B}}(t))} = S(\overline{\rho_{\mathsf{A}}}(t)).
\end{equation}
One way to see how this is the case is by noting that, for fixed R\'enyi index $n$, in the large $N$ approximation,
\begin{equation}
    \overline{\Tr (\rho_{\mathsf{A}}^n(t))} = \Tr (\overline{\rho_{\mathsf{A}}}^n(t)).
\end{equation}
The advantage of connecting the different replica copies of $\rho_{\mathsf{B}}$ is to identify $b_i$ in one copy of the density matrix to $b_i$ in the next. In the end, after evaluating all the sums, there is then a factor 
\begin{equation}
 \left(\frac{1}{N}\right)^n +\left(\frac{N-1}{N}\right)^n \approx 1,
\end{equation}
such that these terms are leading order. The reason that for $\rho_{\mathsf{A}}(t)$ these are subleading corrections is that by identifying $a_i$ in one copy with $a_i$ in the next, the accompanying factor is $\sim \left(\frac{1}{N}\right)^n \approx 0$. 

From a statistical point of view, the above means that at any time, in the large $N$ approximation, the classical statistical ensemble of reduced density matrices $\rho_\mathsf{A}$ is very sharply peaked around its expectation value with the variance being suppressed by factors $\sim 1/N$. 

This feature is thus a consequence of the simplicity of our model, but it is not difficult to see that this still holds if we slightly generalise our model. If we for example consider the Hilbert space decomposition
\begin{equation}
    \cH_{\text{micro}} = \left(\cH_{\mathsf{A},1} \otimes \cH_{\mathsf{B},1}\right)\oplus\left(\ket{0}_{\mathsf{A}} \otimes \cH_{\mathsf{B},2}\right),
\end{equation}
with $|\cH_{\mathsf{A},1}| = N_{\mathsf{A},1}$, $|\cH_{\mathsf{B},1}| = N_{\mathsf{B},1}$ and $|\cH_{\mathsf{B},2}| = N_{\mathsf{B},2}$, then the total Hilbert space dimension is
\begin{equation}
    N = N_{\mathsf{A},1}N_{\mathsf{B},1} + N_{\mathsf{B},2}.
\end{equation}
Under the assumption that $N_{\mathsf{A},1}N_{\mathsf{B},1} \ll N_{\mathsf{B},2}$, then it is not difficult to see that the connected contributions to $\overline{\Tr \rho_{\mathsf{A}}^n}$ will at most be of order
\begin{equation}
    \left(\frac{N_{\mathsf{A},1}N_{\mathsf{B},1}}{N}\right)^n \ll 1.
\end{equation}
Even if we extend the Hilbert space decomposition  further as
\begin{equation}
    \cH_{\text{micro}} = \bigoplus_{E'} \cH_{\mathsf{A};E-E'} \otimes \cH_{\mathsf{B};E'},
\end{equation}
as long as $|\cH_{\mathsf{A};0}|=1$ and 
\begin{equation}
\begin{split}
     |\cH_{\mathsf{A};E-E'}\otimes \cH_{\mathsf{B};E'}|  \begin{cases}
        \ll |\cH_{\mathsf{B};E}| \quad  &E' \neq E\\
        \sim \cO(N) \quad &E'= E,
    \end{cases}
\end{split}
\end{equation}
then the corrections that can come from taking the connected contributions into account similarly vanish in the large $N$ limit.
To recap, for the system $\mathsf{A}$, in the large $N$ approximation, it does not matter whether we average before or after taking the $n$'th power of the partial density matrix: $\Tr{\left(\overline{\rho_{\mathsf{A}}}^n\right)} = \overline{\Tr{\left(\rho_{\mathsf{A}}^n\right)}}$. The connections between the different density matrices are subleading corrections. 

In contrast, for the environment $\mathsf{B}$, 
the story is drastically different. The component that connects the neighbouring density matrices becomes the dominant contribution at late times. 
The disconnected component gives a result analogous to the na\"ive Hawking computation, with a final state where the radiation is maximally mixed. The connected component, analogous to the replica wormholes, purifies the radiation after the Page time, giving a dynamical entropy consistent with unitarity.

\subsection{More realistic models of black hole evaporation}

Our toy model captures both the unitary Page and non-unitary Hawking curves, which was the goal of this paper. If one wished to modify the model to make it a more realistic description of black hole evaporation, there are some properties that one would want to address: 
\begin{enumerate}
    \item Our model has only a single coal/black hole microstate within the microcanonical window.
    \item The Planck scale $l_p$ is not a tunable parameter. Our large $N$ approximation is not a semiclassical approximation, as $\log N$ is the total microcanonical entropy which counts both black hole and radiation microstates.
    \item Our Page time~\eqref{eq:PageTimePurity} is a constant that does not capture the dependence of a realistic black hole's Page time on initial energy or the Planck scale.
\end{enumerate} 
These features are due to the simplicity of our model. There is only one dimensionless input parameter, $N$, while unitary Page curves need at least two: the maximum entropy, and the Page time in Planck units. Adding one more dimensionless parameter, the black hole entropy, to the model would likely be sufficient. These `shortcomings' of our toy model are really a feature, as our goal was to come up with the simplest model that captures both the unitary and non-unitary Page curves and need not capture scales.

\subsubsection*{Larger number of microstates}

If we wished to make the model more realistic, one modification would be to increase the number of black hole/coal microstates from one. A more realistic model would take $e^{S_{\text{BH}}}$ of the $N$ microstates in $\cH_{\text{micro},E}$ to be black hole microstates
so that now
\bne \cH_{\text{micro};E} = (\cH_{\mathsf{A};E}\otimes \ket{0}_{\mathsf{B}}) \oplus (\ket{0}_{\mathsf{A}} \otimes \cH_{\mathsf{B};E}), \ene
where $|\cH_{\text{micro};E}| = N$ as before, but we now have a large number of black hole/coal microstates, with 
\bne N \gg |\cH_{\mathsf{A},E}| = e^{S_\text{BH}} \gg 1. \ene 
This modification increases the maximum entropy of the unitary Page curve from $\log 2$ to $S_{\max} \sim \log  |\cH_{\mathsf{A},E}|$,
the initial black hole microcanonical entropy, but has a negligible effect on the Page time. The physical reason for the absence of an effect on the Page time is that our model still lacks intermediate states between black hole and no black hole; to get a Page time that depends on input parameters, we can introduce intermediate, half-evaporated states, and a notion of locality which captures the property that real black holes evaporate a few Hawking quanta at a time, and not all at once. 

\subsubsection*{Locality}
In the model, as it is right now, there is no notion of locality. To see this, note that we are equally likely to transition to any of the microstates in the microcanonical ensemble that form the basis in which our Hamiltonian is random. In particular, even with the modifications to the microcanonical Hilbert space structure discussed previously, where we increase the initial number of black hole microstates and add intermediate partially evaporated states, with a Hamiltonian that is Gaussian random for the set of states in the microcanonical window~\eqref{eq:micro1}, the most likely transition from the initial state is to the entropically favoured radiation states in $\ket{0}_{\mathsf{A}} \otimes \cH_{\mathsf{B};E}$. 

The eigenvalue statistics of quantum theories of which the classical counterpart is chaotic, are well described by random matrix statistics \cite{Bohigas1984characterization}. Furthermore, within the microcanonical window, the expectation is that the Hamiltonian is unitarily invariant.\footnote{This is in the same spirit as the ETH conjecture: low energy observables are not able to distinguish between closely lying energy eigenstates.} This justified our model's assumption that $H$ is GUE-random in the basis of states for $\cH_{\text{micro};E}$ in~\eqref{eq:micro1}. 
In a theory with local interactions, the Hamiltonian will be of the form $H = H_{\mathsf{A}} + H_{\mathsf{B}} + H_{\text{int}}$, where $H_{\text{int}}$ couples nearby local operators across the shared boundary of the system and environment. 
We have chosen a basis of states that are tensor products of energy eigenstates of the subsystem Hamiltonians $H_{\mathsf{A}}$ and $H_{\mathsf{B}}$. As the interaction between system and environment is turned off, in this basis the Hamiltonian becomes diagonal, so clearly it is a special one.
Locality imposes more structure upon the Hamiltonian than we have assumed and, if we want to capture quantities like the Page time, then we have to account for that.

To model a theory with local interactions, we could make a different ansatz for the Hamiltonian in this basis. As explained above, for a weakly coupled local theory we expect $H$ to be almost diagonal in the $\ket{E'}_{\mathsf{A}} \otimes \ket{E-E'}_{\mathsf{B}}$ basis. 
Instead of choosing the GUE, we could think of opting for a probability distribution like
\begin{equation}
    \mu(H) \propto \exp\left(-\sum_{i,j} |H_{ij}|^2 c_{ij}\right), \qquad c_{ij} \sim e^{(i-j)^2\alpha},
\end{equation}
with $\alpha > 0$. In that case, if $i$ and $j$ are far from each other and $H_{ij}$ is not very small, this Hamiltonian is heavily suppressed. This probability density function is by design no longer invariant under unitary transformations. Therefore, after the decomposition of the Hamiltonian into unitary matrices and eigenvalues, there will be a complicated measure over the unitary matrices alongside the Vandermonde determinant for the eigenvalues. As a consequence, the computation is much more complicated.
In line with the ETH conjecture, we want to ensure that whatever measure we pick gives approximate unitary invariance in narrow energy bands with many states.

Another possibility more closely related to the GUE is that the probability density function will be dependent on a trace like in the GUE, but of the commutator of some fixed diagonal matrix $M$ with $H$:
\begin{equation}
    \mu(H) \propto \exp \left(- \Tr \left([M,H]^2\right)\right) = \exp \left(- \sum_{i,j} (m_i-m_j)^2 |H_{ij}|^2\right).
\end{equation}
Yet another possibility is to manually add locality in the interaction Hamiltonian by giving it a small support only on the interaction Hilbert space:
\begin{equation}
    \cH = \bigoplus_{E_1, E_2,E_3} \cH_{\mathsf{A'};E-E_1} \otimes \cH_{\mathsf{A}_{\text{int}};E_2} \otimes \cH_{\mathsf{B}_{\text{int}};E_3} \otimes \cH_{\mathsf{B'};E_1-E_2-E_3}.
\end{equation}
The interaction Hilbert space is given by 
\begin{equation}
    \cH_{\text{int}}=\bigoplus_{E_2,E_3} \cH_{\mathsf{A}_{\text{int}};E_2}\otimes \cH_{\mathsf{B}_{\text{int}};E_3}\, ,
\end{equation}
and we take $|\cH_{\text{int}}|$ to be small. Furthermore, we take $|\cH_{\mathsf{B}';E}|$ to be much larger than any other subspace.
The reason we decompose the interaction Hilbert space into an $\mathsf{A}_{\text{int}}$- and $\mathsf{B}_{\text{int}}$-part is because we still want to be able to compute the entropies in the entire $\mathsf{A}$ and $\mathsf{B}$, which are the primed parts and interaction parts together. 

The expectation is that the behaviour of the system will be similar to what was found here since moving all the energy into the environment is entropically favoured. However, the exact time dependence and the maximum of the entropies might change and take on dependence on the sizes of the subspaces.

 \subsection*{Planckian and stringy corrections}
 
Our model has one dimensionless parameter $N$, the number of states in the microcanonical window.
If we wished to, how would we incorporate stringy or Planckian corrections?
There are several other dimensionless parameters we could add like the number of black hole states $e^{S_{\text{BH}}}$ and coefficients of higher order terms in the matrix model potential.
Non-perturbative gravitational corrections of the form $e^{-1/G_N}$ would appear as corrections in this new parameter.  
Part of the dynamics of a more realistic model could
 be accounted for by including higher-order terms in the matrix
model potential, and the precise interpretation of these terms
would depend on the way the toy model is embedded in a more
realistic set-up.

\subsection{Information transfer}
\label{sec: Information transfer}
Now we will study how the evaporation discussed in the toy model transfers information from the black hole to the radiation, similarly to the Hayden-Preskill process \cite{Hayden2007mirror}. In the original process, Alice throws a diary into a black hole. This diary is maximally entangled with some sets of qubits in the hand of Charlie. While the black hole evaporates, Bob collects the Hawking radiation. In this process, Bob's final state is maximally entangled with Charlie's system, showing how the information has been transferred from the black hole to the Hawking radiation. 

We will study a different process, as our evaporation is not imposed by moving qubits from the black hole Hilbert space to the radiation Hilbert space at discrete time steps but rather by assuming chaotic dynamics within a microcanonical energy window. The model will also differ in another aspect. In the Hayden-Preskill process, Alice throws her diary into an existing (either young or old) black hole where Bob has collected the radiation that has already come out of the black hole. Here Bob's radiation will be maximally entangled with the remainder of the black hole. 

Our model is more analogous to a burning piece of coal (or a burning diary). We let Alice prepare two copies of the diary and maximally entangle them. Then she will burn one of the diaries in her system while giving the other diary to Charlie. We will take Bob's initial system, the environment, to be unentangled with the burning diary, but he collects the radiation coming off of the diary as it burns. We want to track where the information in the burning diary is as a function of time. When the diary has evaporated entirely, the information in it must have moved into the radiation. To quantify the information of the burning diary, we will compute the mutual information between the intact diary in Charlie's possession and the separate subsystems of Alice and Bob.

The effective Hilbert space is taken to be the tensor product between Charlie's Hilbert space $\cH_{\mathsf{C}}$ and $\cH_{\mathsf{AB};E}$; the combined microcanonical Hilbert space of Alice and Bob:
\begin{equation}
\begin{split}
     \cH_{\text{eff}} 
     = \cH_{\mathsf{AB};E}\otimes \cH_{\mathsf{C}}, \qquad \cH_{\mathsf{AB};E}=\left(\cH_{\mathsf{A};E}\otimes \ket{0}_{\mathsf{B}}\right)\oplus \left(\ket{0}_{\mathsf{A}}\otimes \cH_{\mathsf{B};E}\right).
\end{split}
\end{equation}
Here we take both the microcanonical Hilbert space $\cH_{\mathsf{A};E}$ and $\cH_{\mathsf{C}}$ to be $m$-dimensional and $\cH_{\mathsf{B};E}$ to be $(N-m)$-dimensional, 
\begin{equation}
    \cH_{\mathsf{A};E} = \bigoplus_{i=1}^m \ket{a_i}, \qquad \cH_{\mathsf{B};E} = \bigoplus_{i=1}^{N-m} \ket{b_i}, \qquad \cH_{\mathsf{C}} = \bigoplus_{i=1}^m \ket{c_i}.
\end{equation}
We thus have $|\cH_{\mathsf{AB};E}|=N$, and $|\cH_{\text{eff}}|=mN$. Furthermore, we assume $N \gg m$. 
We will take our initial density matrix to be of the form
\begin{equation}
    \rho_0 = \rho_{\mathsf{AC}}\otimes \rho_{\mathsf{B}}.
\end{equation}
Here $\rho_{\mathsf{AC}}$ is a maximally entangled state between the diary in $\mathsf{A}$ and the auxiliary system in $\mathsf{C}$, and we will take the environment to be initially in the ground state: 
\begin{equation}
    \rho_{\mathsf{AC}} = \ket{\psi_{\mathsf{AC}}}\bra{\psi_{\mathsf{AC}}}, \qquad \ket{\psi_{\mathsf{AC}}} = \frac{1}{\sqrt{m}}\sum_{i=1}^m \ket{a_i\, c_i}, \qquad \rho_{\mathsf{B}} = \ket{0}\bra{0}.
\end{equation}
We will take the Hamiltonian to act trivially on $\cH_{\mathsf{C}}$ and randomly on the $N$-dimensional subspace $\cH_{\mathsf{AB};E}$:
\begin{equation}
    H = H_{\text{GUE}(N)} \otimes \mathds{1}_{\mathsf{C}}.
\end{equation}
That is, to model the chaotic evaporation of the diary in $\cH_{\mathsf{AB};E}$, we take the Hamiltonian to be drawn randomly from the $N \times N$ Gaussian unitary ensemble (GUE). Based on the computation in Sec. \ref{sec: toy model}, the expectation is that now the diary will evaporate as time evolves, and the information in it will move into the radiation in the environment.
The mutual information between the separate systems is defined as
\begin{equation}
I(\rho_{\mathsf{X}},\rho_{\mathsf{Y}},\rho_{\mathsf{XY}}) \coloneqq S(\rho_{\mathsf{X}}) + S(\rho_{\mathsf{Y}}) - S(\rho_{\mathsf{XY}}), \qquad
    S(\rho_{\mathsf{X}}) \coloneqq - \partial_n \Tr \rho_{\mathsf{X}}^n \bigg \rvert_{n=1}.
\end{equation}
This quantity measures the reduction in uncertainty about the state in $\mathsf{X}$ after measuring the state in $\mathsf{Y}$, and vice versa.

\subsubsection{Information in the system}
 First, we want to compute the mutual information between the system $\mathsf{A}$ and the auxiliary system $\mathsf{C}$. In order to compute this, we need the following quantities:
\begin{equation}
    \begin{split}
         \overline{\Tr (\rho_{\mathsf{A}}^n)} &=\frac{\overline{g}^n}{m^{n-1}}+(1-\overline{g})^n , \qquad
         \overline{\Tr( \rho_{\mathsf{AC}}^n)}= \frac{(1-\overline{g})^n}{m^{n-1}} + \overline{g}^n,\qquad
        \overline{\Tr( \rho_{\mathsf{C}}^n)} = \frac{1}{m^{n-1}}.
        \label{eq: purities black hole hayden preskill}
    \end{split}
\end{equation}
The computation of these quantities is done in a similar manner as in Sec. \ref{sec: toy model}, and we have again made the approximation $\overline{g^n} \approx \overline{g}^n$. We then find\footnote{As before, we are computing the von Neumann entropy by averaging before taking the derivative instead of the other way around.} 
\begin{equation}
\begin{split}
     \overline{S(\rho_{\mathsf{A}})} &= \overline{g} \log m - \overline{g} \log \overline{g}- (1-\overline{g})\log (1-\overline{g}), \\
    \overline{S(\rho_{\mathsf{C}})} &= \log m,\\
  \overline{S(\rho_{\mathsf{AC}})} &= (1-\overline{g}) \log m -  \overline{g} \log \overline{g} - (1-\overline{g}) \log (1-\overline{g}).
\end{split}
\end{equation}
From here, the mutual information can be computed to be
\begin{equation}
\overline{I(\rho_{\mathsf{A}},\rho_{\mathsf{C}},\rho_{\mathsf{AC}})} = 2\overline{g}\log m.
\end{equation}
Initially, the mutual information between $\mathsf{A}$ and $\mathsf{C}$ is given by $2 \log m$. 
At late times, it goes to 0. This means that, as expected, the information leaves the system $\mathsf{A}$ as time evolves. 

\subsubsection{Information in the environment}
\begin{figure}
    \centering
    \includegraphics[width=0.7\textwidth]{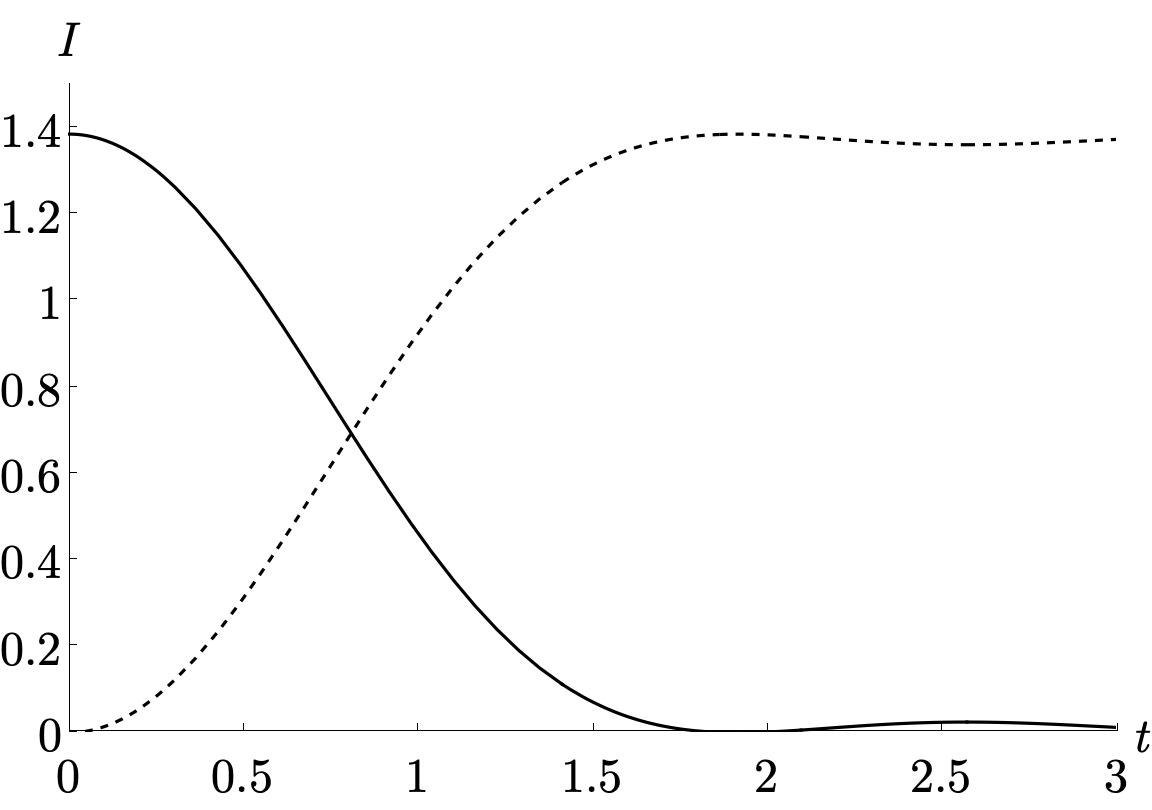}
    \caption{The mutual information $I$ between the auxiliary system $\mathsf{C}$ and various systems is shown for $m=2$. The solid line is the mutual information between $\mathsf{A}$ and $\mathsf{C}$, the dashed line is the mutual information between $\mathsf{B}$ and $\mathsf{C}$. It can be seen that the mutual information between $\mathsf{A}$ and $\mathsf{C}$ decreases from $2 \log m$ at $t=0$ to zero at late times, while the mutual information between $\mathsf{B}$ and $\mathsf{C}$ increases from 0 initially to $2 \log m$ at late times. This is the average mutual information computed for 100 Hamiltonians drawn from the GUE with $N=100$.}
    \label{fig: mutual info new}
\end{figure}

Similarly, we can compute the mutual information between the environment $\mathsf{B}$ and the diary in $\mathsf{C}$. Computing the relevant quantities gives
\begin{equation}
    \begin{split}
        \overline{ \Tr( \rho_{\mathsf{B}}^n)} &=\frac{(1-\overline{g})^n}{m^{n-1}}+\overline{g}^n,\qquad
        \overline{ \Tr (\rho_{\mathsf{BC}}^n)} = \frac{\overline{g}^n}{m^{n-1}} + (1-\overline{g})^n.
         \label{eq: purities radiation hayden preskill}
    \end{split}
\end{equation}
Note that these are exactly the quantities one would expect from unitarity. These give the entropies
\begin{equation}
    \begin{split}
    \overline{S(\rho_{\mathsf{B}})} &= -\overline{g}\log \overline{g} - (1-\overline{g})\log (1-\overline{g}) + (1-\overline{g})\log m,\\
    \overline{S(\rho_{\mathsf{BC}})} &= -\overline{g}\log \overline{g} - (1-\overline{g})\log (1-\overline{g}) + \overline{g}\log m.
    \end{split}
\end{equation}
The mutual information between the environment and the intact diary in $\mathsf{C}$ is thus given by
\begin{equation}
\overline{I(\rho_{\mathsf{B}},\rho_{\mathsf{C}},\rho_{\mathsf{BC}})} = 2(1-\overline{g})\log m.
\end{equation}
From here it is clear that the information gradually moves into the environment as time evolves. The time evolution of the mutual information is visualised in Fig. \ref{fig: mutual info new} for $m=2$. 

It is interesting to note what we would have obtained if we did not take into account the replica wormholes. The equations in \eqref{eq: purities black hole hayden preskill} will be unaltered when first averaging and then taking the $n$'th power. However, the ones in \eqref{eq: purities radiation hayden preskill} will differ:
\begin{equation}
    \begin{split}
         \Tr (\overline{\rho_{\mathsf{B}}}^n) &=\overline{g}^n+ \frac{(1-\overline{g})^n}{N^{n-1}},\qquad
         \Tr (\overline{\rho_{\mathsf{BC}}}^n) = \frac{\overline{g}^n}{m^{n-1}} + \frac{(1-\overline{g})^n}{(mN)^{n-1}}.
         \label{eq: purities radiation hayden preskill 2}
    \end{split}
\end{equation}
This gives the entropies
\begin{equation}
\begin{split}
    S(\overline{\rho_{\mathsf{B}}}) &= -\overline{g} \log \overline{g} - (1-\overline{g}) \log (1-\overline{g}) + (1-\overline{g})\log N,\\
     S(\overline{\rho_{\mathsf{BC}}})&= - \overline{g}\log \overline{g} - (1-\overline{g}) \log (1-\overline{g})  - (1-\overline{g})\log N+ \log m.
\end{split}
\end{equation}
This means that without taking into account the replica wormholes, mutual information between the radiation and the diary is
\begin{equation}
I(\overline{\rho_{\mathsf{B}}},\overline{\rho_{\mathsf{C}}},\overline{\rho_{\mathsf{BC}}}) = 0.
\end{equation}
Na\"ively, the information does not show up in the radiation at all.

We have thus seen how the information in a chaotically burning diary shows up in the radiation. Modelling the chaos is done by taking the time evolution of the diary and radiation to be drawn randomly from an ensemble of the GUE. 
Using the replica trick, we have computed the average von Neumann entropies and used those to quantify the mutual information between the intact diary in $\mathsf{C}$ and either the system with the burning diary $\mathsf{A}$ or the environment with the radiation $\mathsf{B}$. The information enters the environment through the replica wormholes: if one does not take the connected components into account, the mutual information between the environment and the diary in $\mathsf{C}$ is zero at all times. 

\section{Discussion}
\label{sec: conclusion and discussion}
In this paper, we have shown how the non-unitary Hawking curve and the unitary Page curve naturally appear in a simple toy model, assuming ergodicity, energy conservation and a gapped system. By observing the state of the environment $\mathsf{B}$ in the large $N$ approximation, at late times one finds a maximally mixed density matrix for which the von Neumann entropy grows to $\log N$. This is our analogue of the thermal Hawking radiation, leading to the Hawking curve for the von Neumann entropy.

However, by more carefully probing the entropy through the replica trick and computing the quantities $\overline{\Tr \rho_{\mathsf{B}}^n}$ in the large $N$ approximation, we have found a curve for the entropy consistent with unitarity, our analogue of the Page curve. Studying the contraction patterns of the density matrices, one finds two leading order contributions. The first one is a disconnected component, and if only taking this one into account, the result reduces to the Hawking curve. However, the second leading contribution cyclically connects the neighbouring density matrices, forming an analogue of the replica wormholes. This contribution starts to dominate at the Page time. By taking this term into account, the von Neumann entropy decays after the Page time.
Our model clarifies how semi-classical replica wormholes are able uncover a Page curve consistent with microscopic unitarity. Gravitational computations involve some coarse graining but the coarse graining is such that they map pure initial states into classical statistical mixtures of pure states at later times. Semi-classical replica computations can diagnose whether the statistical mixture consists of pure states or not, but are unable to accurately identify the indidual pure states themselves. 
Our Hamiltonian ensembles are unitary-invariant. The relation between unitary matrix contractions and replica wormholes is a consequence of the Haar averaging. The eigenvalue distribution is important for the shapes of the radiation entropy curves $\overline{S(\rho_\mathsf{B})}$ and $S(\overline{\rho_\mathsf{B}})$, through the averaged spectral form factor $\overline{g}(t)$. A $\overline{g}(t)$ that returns to order $1$ too soon, as may be expected in systems such as the simple harmonic oscillator, would give shapes that are inconsistent with the unitary and non-unitary Hawking radiation entropy curves. 

Previous work on time evolution with random Hamiltonians from the GUE shows that while the GUE is a good approximation for late time physics of local quantum chaotic theories, at early times there is a discrepancy \cite{Cotler2017chaos, Vijay2018finite}. The GUE is unitarily invariant, which makes computations tractable, but as a consequence, it only describes non-local physics. These theories scramble much faster than is to be expected for local theories of quantum chaos. However, at late times, all local operators in quantum chaotic systems have had time to spread and the notion of locality is lost and the useful information of the system is found in the spectrum. The GUE is a good description for the spectral properties of the Hamiltonian \cite{Bohigas1984characterization}, which means the GUE is capable of accurately describing the late time physics. Even though the GUE thus fails to capture local physics at early times, the late-time conclusions of this work remain even for local chaotic systems. At late times, it is clear that the leading saddle is the connected contribution as in Fig. \ref{fig: connected rho}, whereas only considering the disconnected components will give an entropy $\sim \log N$. Initially, when the state is exactly pure in $\mathsf{A}$ and $\mathsf{B}$ separately, the disconnected component as in Fig. \ref{fig: disconnected rho} will be the only contribution. Then at some point there will be an exchange in dominance between these two saddles. Therefore, the qualitative picture will not change; only the time-scales and the maximum of the entropy might be affected by introducing local physics. In particular, the recovery of a unitary Page curve only requires fairly general input; assuming a microcanonical window with chaotic dynamics is sufficient to find a unitary Page curve once the replica wormholes are included.

\subsection{Comparison to other work}

\subsubsection*{Toy models with dynamical unitary Page curves}
There are many toy qubit models of black hole evaporation, see~\cite{Mathur:2011wg} and references therein, and quenched field theory models~\cite{Dadras:2020xfl,Zhang:2023gzl}. The difficult aspect of any toy model is capturing both the non-unitarity of semiclassical black holes and showing how unitarity is restored.

An approach that is close to our own is looking for dynamical Page curves and replica wormholes in SYK-based models~\cite{Su:2020quk,Wang:2023vkq}. It is close in part because their Hamiltonians are also drawn from an ensemble, and because of the attempt to capture replica wormholes, but it is different because of the specificity of SYK. 

There have also been random unitary circuit models of black hole evaporation~\cite{Piroli:2020dlx}, and a model that attempts to incorporate energy conservation with Haar-random unitary evolution~\cite{Yoshida:2018ybz}.
The ``operator-gas approach'' \cite{Liu2021dynamical} finds a dynamical mechanism for the Page curve from quantum chaos without resorting to typicality, by studying the support of operators and modelling the probability of creating a void in this support. 
\cite{Liu2021equilibrated} computes the late-time entropy by projecting the states onto diagonal subspaces called equilibrated states. \cite{Krishnan:2021faa} models the black hole as a box leaking a gas of chaotic hard spheres.
These models dynamically capture the unitary Page curve but their underlying mechanisms are fundamentally different from ours. None evolve within microcanonical windows with an ensemble of Hamiltonians.

\subsubsection*{Page's theorem and the West Coast model}
In the original papers on the information paradox by Page \cite{Page1993average, Page1993information}, the main assumption is that the total system is in a random pure state in the factorised Hilbert space $\cH = \cH_{\mathsf{A}} \otimes \cH_{\mathsf{B}}$. By averaging over this random state with the Haar measure, the average entropy in the smaller subsystem is computed. Increasing the dimension of this smaller subsystem step-by-step, it can be found that the von Neumann entropy follows the well-known Page curve. The authors of \cite{Penington2022replica} find a unitary Page curve from a toy model with random Hamiltonians. Besides time evolution with a random Hamiltonian, a set of random states is considered, similarly as in \cite{Page1993average}. The average overlap between these states is what gives rise to the replica wormholes in this computation. Again, by manually increasing the dimension of the radiation Hilbert space, the entropy is shown to go up and then down again. Both \cite{Page1993average} and \cite{Penington2022replica} thus take the state to be a random pure state, modelling uncertainty in the exact microscopic details of the state. Assuming evaporation, the dimensionality of the radiation Hilbert space then grows and this enforces the unitary Page curve.

The main difference with our approach is that we instead take the Hamiltonian to be random in the microcanonical window, motivated by chaos RMT and modelling uncertainty in microscopic physics. Without assuming evaporation, we show how replica wormholes are able to restore unitarity in chaotic systems.

\subsubsection*{Entanglement and random dynamics}
Time evolution with GUE-random Hamiltonians, and the resulting entanglement dynamics, have been previously studied in the condensed matter community~\cite{Znidarivc2012subsystem,Cotler2017chaos, Vijay2018finite,Chernowitz2021entanglement,You2018entanglement}. One important result is that the smaller subsystem evolves to the maximally mixed state. 

To compare, we take the Hamiltonian to be random only within microcanonical windows, to model energy conservation, and this ensures that the entanglement entropy is small at late times due to the argument outlined in Sec. \ref{sec: toy model}.

\subsection{Future work}

\subsubsection*{Extending the model}
One possible direction for future work is to extend the existing model. For example, the model in Sec. \ref{sec: Information transfer} might be extended to more closely resemble the actual Hayden-Preskill model, where the diary is thrown into an existing black hole and where Bob has already collected all of the early radiation. Another future direction could be to study the higher moments of the spectral form factor. The computation of the von Neumann entropy assumes the spectral form factor is self-averaging, which is only true before the dip time. After the dip time, the assumption $\overline{g^n}\approx \overline{g}^n$ is not valid anymore and we need additional information to be able to compute the $n\to 1$ limit of the R\'enyi entropies.
The dip time is of order $\sqrt{N}$, so for large $N$ this is at very late times. At these late times, $\overline{g^n}$ can grow close to 1 and thus give rise to large R\'enyi entropies. In any case, a better estimate for the late time value of the higher moments of the spectral form factor would be valuable for better understanding chaotic systems at late times.

\subsubsection*{(B)CFT models}
In future work, we would like to return to one of our motivating questions: what are the necessary and sufficient conditions on a holographic (B)CFT to get a unitary Page curve? To investigate this, we need to upgrade our model. 
We could take a single holographic CFT, and consider the evolution of a pure high energy primary state. If the CFT's OPE coefficients are not exact, but drawn from an ensemble, such as in~\cite{Belin2021,Belin2022generalized, Belin2022non, Anous2022ope, Belin2023approximate}, can the von Neumann entropy of the averaged state still be zero at all times? At first sight the answer would seem to be yes, because this ensemble will map an initial pure state into a classical statistical mixture of pure states, just as in our toy model. To what extent can we introduce non-unitarities in the operator dimensions and OPE coefficients and still get a unitary Page curve? What is the set of CPTP maps that give the same unitary Page curve as ordinary time evolution? We can ask these questions in bottom-up or specific top-down holographic CFTs models, singly or doubly holographic, with or without radiation baths. For other work connecting statistical properties of CFTs with gravity, see~\cite{Altland2021late, Chandra2022semiclassical, Cotler2021AdS3}.
\subsubsection*{Eigenstate thermalisation hypothesis}
The eigenstate thermalisation hypothesis (ETH) for the matrix elements of simple operators in the energy eigenbasis is an ansatz that leads to thermalising behaviour in quantum systems. Because of ETH's connection to chaotic systems and random matrix theory, it is natural to ask whether it alone could capture both the non-unitary Page curve and replica wormhole-like unitarity-restoring corrections. One toy model would be to take the interaction Hamiltonian between system and environment to be $H_{\text{int}} = \lambda \cO_{\mathsf{A}}\otimes \cO_\mathsf{B}$, a coupling between simple local operators, with one or both of the operators satisfying the ETH ansatz.
\subsubsection*{Non-Markovian open quantum systems}
We would like to connect our unitarity-restoring random matrix contractions with non-Markovian corrections to the Lindblad master equation. The radiation subsystem is an open quantum system, and Lindbladian evolution is a good approximation when the dynamics are Markovian. If the Lindblad operators are Hermitian, then the purity of the state decreases monotonically and the long-time average is maximally mixed~\cite{Manzano_2020}. Markovian dynamics are inconsistent with a unitary Page curve, and it would be interesting to see how and when non-Markovian corrections appear in our present and future models. One approach may be to extend the Lindblad equation to the dynamics of replicated open systems.

\subsubsection*{Factorisation problem}
At the heart of both the factorisation problem and the replica wormhole story are Euclidean saddles that connect copies of gravitational systems. The factorisation problem is a tension between the calculation of $Z(\beta)^n$ in gravitational theories, which in some examples does not factorise due to connected geometries in the gravitational path integral, and the factorisation predicted from the boundary dual being a single CFT theory. The wormholes that connect multiple boundaries have a statistical interpretation, they quantify the correlations in the classical statistical ensemble which corresponds to semi-classical gravity
\cite{Belin2021,Belin2022generalized,Belin2022non,Anous2022ope,Belin2023approximate,ignorance,Pollack2020eigenstate}. 
The contribution of the wormholes can heuristically be expressed as
\begin{equation}
    \int dH \mu(H) \left(\Tr e^{-\beta H}\right)^n  - \left(\int dH \mu(H) \Tr e^{-\beta H}\right)^n,
    \label{eq: Euclidean wormholes}
\end{equation}
where one integrates over all Hamiltonians which are semi-classically indistinguishable. 
It is interesting to notice that our toy model strongly suggests that replica wormholes appear to have a similar statistical origin. It would be interesting to pursue this analogy further and to explore to what extent the various issues associated with multi-boundary wormholes have a replica wormhole counterpart. 

\acknowledgments
We thank Ramesh Ammanamanchi, Luis Apolo, Igal Arav, Vladimir Gritsev, Diego Li\v ska, Dominik Neuenfeld, Boris Post, Mark van Raamsdonk, Kamran Salehi Vaziri for useful discussions.

The work of JH is supported by the Dutch Black Hole Consortium with project number NWA.1292.19.202 of the research programme NWA which is (partly) financed by the Dutch Research Council (NWO). The work of AR is supported by the Stichting Nederlandse
Wetenschappelijk Onderzoek Instituten (NWO-I) through the Scanning New Horizons
project, by the Uitvoeringsinstituut Werknemersverzekeringen (UWV), by FWO-Vlaanderen project G012222N, and by the Vrije Universiteit Brussel through the Strategic Research Program High-Energy Physics. JdB is supported by the European Research Council under the European Union's Seventh Framework Programme (FP7/2007-2013), ERC Grant agreement ADG 834878.

\appendix
\section{Entanglement spectrum probability distribution} \label{app:spectrumPDF}
The entanglement spectrum of our time evolved state, for a given draw of a random Hamiltonian, is specified by a single real number
\bne |\lambda|^2 = (e^{iHt})_{11} (e^{-iHt})_{11}, \ene
where $H$ is a random matrix in the GUE ensemble; this simplicity of the entanglement spectrum is manifest in the reduced matrix on ${\mathsf{A}}$ which can be written as
\bne \rho_{\mathsf{A}}(t) = |\lambda|^2 \rho_{\mathsf{A}} (0) + (1-|\lambda|^2) \rho_{{\mathsf{A}},\text{vac.}}\, . \ene
The Renyi entropies of $\mathsf{A}$ and $\mathsf{B}$ are simple functions of $|\lambda|^2$, so if we knew the probability distribution function (PDF) of $|\lambda|^2$ then we could calculate quantities like $\overbar{S_{\mathsf{B}}^{(n)}(t)}$ directly by integrating over the PDF.

The PDF we want is 
\bne \begin{split} \label{eq:PDF} P(|\lambda|^2) &= \int d\mu_{\text{GUE}} (H) \delta \left( |\lambda|^2 - (e^{iHt})_{11} (e^{-iHt})_{11}\right) \\
&= \int d\mu (\vec{\lambda}) d\mu (U) \delta \left( |\lambda|^2 - \left|\sum_i |U_{1i}|^2 e^{i\lambda_i t}\right|^2 \right).
\end{split} \ene
In the second line, we have split the integral: $\mu (U)$ is the Haar measure on $U(N)$ and $\mu (\lambda)$ is the joint PDF on eigenvalues of $H$. 

While we can explicitly evaluate~\eqref{eq:PDF} for low values of $N$, we were not able to for general $N$ and we have soft evidence that the problem isn't tractable: if we consider the apparently easier calculation that ignores the unitary matrix elements and integrals in~\eqref{eq:PDF}, then we are calculating 
\bne P\left(\left|\sum_i  e^{i\lambda_i t}\right|^2\right).\ene
We recognise the argument of this PDF as the spectral form factor (SFF). To our knowledge, the PDF of the SFF for the GUE, or other standard random ensembles, is not known; the mean and the variance are known~\cite{Liu:2018hlr}, but not the full PDF. It would be of interest to the RMT community to pursue the calculation of the PDF of the SFF for Gaussian ensembles further.

\section{Unitary integrals and Weingarten functions}
\label{app:weingarten}
The unitary integrals can be evaluated using the Weingarten functions:
\begin{equation}
    \begin{split}
        \int_{U_N} U_{i_1 j_1} \dots U_{i_q j_q}U^*_{i_1' j_1'} \dots U_{i_q' j_q'}^*= \sum_{\sigma, \tau \in S_q} \delta_{i_1 i_{\sigma(1)}'} \dots \delta_{i_q i'_{\sigma(q)}} \delta_{j_1 j_{\tau(1)}'} \dots \delta_{j_q j_{\tau(q)}'} \mathsf{Wg}(\sigma \tau^{-1},N).
        \label{eq: unitary contraction weingarten}
    \end{split}
\end{equation}
The Weingarten function $\Wg(\sigma \tau^{-1}, N)$ only depends on the conjugacy class of $\sigma \tau^{-1}$ and $N$. We have listed below the values for $1 \leq q \leq 4$ \cite{Mironov1996}:
\begin{equation}
    \begin{split}
        q = 1: \qquad &\Wg([1],N) = \frac{1}{N},\\
        q=2: \qquad &\Wg([1,1],N) = \frac{1}{N^2-1}, \\ &\Wg([2],N) = \frac{-1}{N(N^2-1)},\\
        q=3: \qquad &\Wg([1,1,1],N) = \frac{N^2-2}{N(N^2-1)(N^2-4)}, \\
        &\Wg([2,1],N) = \frac{-1}{(N^2-1)(N^2-4)}, \\
         &\Wg([3],N) = \frac{2}{N(N^2-1)(N^2-4)},\\
         q=4: \qquad &\Wg([1,1,1,1],N) = \frac{N^4-8N^2+6}{N^2(N^2-1)(N^2-4)(N^2-9)},\\
         &\Wg([2,1,1],N) = \frac{-1}{N(N^2-1)(N^2-9)},\\
         &\Wg([2,2],N) = \frac{N^2+6}{N^2(N^2-1)(N^2-4)(N^2-9)},\\
         &\Wg([3,1],N) = \frac{2N^2-3}{N^2(N^2-1)(N^2-4)(N^2-9)},\\
         &\Wg([4],N) = \frac{-5}{N(N^2-1)(N^2-4)(N^2-9)}.
    \end{split}
\end{equation}
For large $N$, the Weingarten functions have the asymptotic expression 
\begin{equation}
    \Wg(\sigma \tau^{-1},N) = N^{-q-|\sigma \tau^{-1}|}\prod_i \left((-1)^{C_i-1}c_{C_i-1}\right) + O(N^{-q-|\sigma \tau^{-1}|-2}).
    \label{eq: asymptotic Weingarten}
\end{equation}
Here $|\sigma \tau^{-1}|$ is the minimal number of transpositions needed to write $\sigma \tau^{-1}$, $\sigma \tau^{-1}$ contains cycles of length $C_i$ and $c_n$ is a Catalan number:
\begin{equation}
    c_n = \frac{(2n)!}{n! (n+1)!}.
\end{equation}

\section{Computation: Higher R\'enyi entropies}
\label{app: higher renyi}

We will work in the large $N$ approximation. We will consider fixed $n$, such that combinatorial factors cannot cancel the $1/N$ suppression. We will compute the higher R\'enyi entropies, defined as
\begin{equation}
    \overline{S^{(n)}(\rho_{\mathsf{B}})} = \frac{\log \overline{\Tr \left(\rho_{\mathsf{B}}^n\right)}}{1-n}.
\end{equation}
Note that we take the average before computing the logarithm. We are thus computing the annealed instead of the quenched R\'enyi entropies.
The expression for $\Tr \left(\rho_{\mathsf{B}}^n\right)$ reads 
\begin{equation}
    \begin{split}
    \Tr \left(\rho_{\mathsf{B}}^n\right)&= 
    \sum_{k_1=1}^N \dots \sum_{k_{2n}=1}^N ~ 
    \sum_{b_1=1}^N \dots \sum_{b_n=1}^N 
    \left(\prod_{m=1}^n \delta_{1b_m} + \prod_{m=1}^n (1-\delta_{1b_m})\right)\times\\
  &\phantom{=}\times \prod_{i=1}^n  
   e^{i(\lambda_{k_{2i-1}}-\lambda_{k_{2i}})t} U_{b_i k_{2i-1}} U_{k_{2i-1} 1}^\dagger U_{1 k_{2i}}  U_{k_{2i} b_{i+1}}^\dagger,
    \label{eq: trace n 2}  
    \end{split}
\end{equation}
where $b_{n+1} = b_1$.
We will compute the leading order contribution to the R\'enyi entropies for both the first term, with $b_i = 1$ for all $i$, and the second term, with $2 \leq b_i \leq N$ for all $i$, separately. It will be useful to consider what the maximal contribution from the separate components is.
The maximum contribution from the Weingarten functions is $\sim N^{-2n}$,\footnote{By the symbol $\sim$ we mean here that this is the leading order in $N$.} where we used \eqref{eq: asymptotic Weingarten} with $q = 2n$ and $|\sigma \tau^{-1}| \geq 0$. The maximal contribution from the $b_i$'s is when they are all contracted to themselves, giving a contribution $\sim N^n$. The maximal contribution from the $2n$ different $k_i$'s is when they are all contracted to themselves, giving $\sim W^{n} \sim N^{2n}$. 

\subsection{\texorpdfstring{$b_i=1$}{bi=1}}
For the first term in \eqref{eq: trace n 2} all $b_i$ are set equal to 1. 
Since the Weingarten function is of order $N^{-2n}$ or lower, we need a contribution of at least order $\sim N^{2n}$ from the $k_i$'s for the contraction to survive in the large $N$ limit. There is exactly one $k_i$-contraction pattern that gives this contribution, and that is when all the $k_i$'s are contracted to themselves. 
Writing the unitaries in the notation of \eqref{eq: unitary contraction weingarten}, we have 
\begin{equation}
    \begin{split}
        U_{b_i k_{2i-1}} &= U_{i_{2i-1} j_{2i-1}}~, \qquad 
        U_{1 k_{2i-1}}^* = U_{i'_{2i-1} j'_{2i-1}}^*~,\\
        U_{1 k_{2i }} &= U_{i_{2i} j_{2i}}~,  ~\,\quad \qquad
        U_{b_{i+1} k_{2i} }^* = U_{i'_{2i} j'_{2i}}^*~,
\label{eq: contractions weingarten}
    \end{split}
\end{equation}
where we set $b_i = b_{i+1} = 1$ here.
The only contraction pattern that survives the large $N$ limit therefore is the one with $\tau = e$. In order to not pick up any further suppression from the Weingarten function, we need $\sigma \tau^{-1} = e$ and therefore the only possibility is $\sigma = e$. The corresponding contribution is, in the large $N$ approximation,
\begin{equation}
    \overline{\Tr \left(\rho_{\mathsf{B}}^n\right)}\bigg \rvert_{b_i=1} =  \overline{g^n}.
\end{equation}

\subsection{\texorpdfstring{$b_i = 2, \dots, N$}{bi=2,...,N}}
In this case, any of the $b_i$'s can take $N-1$ values. The maximum contribution this gives is or order $\sim N^{n}$, when every $b_i$ is contracted with itself. 
The maximum contribution from the $k_i$'s however, is no longer $\sim N^{2n}$; we cannot contract any $k_i$ with itself without making a cross-contraction. As can be seen from \eqref{eq: contractions weingarten}, without a cross-contraction the corresponding $b$ would be contracted with 1. This is the one value $b$ is not allowed to have for the second term in \eqref{eq: trace n}.
We thus have to take cross-contractions, where each cross-contraction involves a suppression by at least one factor of $N$. Then we would get a contribution (from two pairs of $k$'s contracted to themselves) of order at most $\sim W/N = gN$, which at early times is of order $N$. 
If we don't contract a $k_i$ with itself but pair it with another $k_j$, together it will contribute a factor $N$ to the sum.  
Therefore, the maximum contribution from all $k$'s together can be of order $N^{n}$. We can conclude that we need all of the $b_i$'s to be contracted with themselves to be able to remain leading order. 

Let us now examine the suppression of one of these cross-contractions. Writing again the unitaries in the notation of \eqref{eq: unitary contraction weingarten}, we have 
\begin{equation}
    \begin{split}
        U_{1 k_{2i }} &= U_{i_{2i} j_{2i}}~,  \qquad \qquad 
        U_{b_{i+1} k_{2i} }^* = U_{i'_{2i} j'_{2i}}^*~,\\
        U_{b_{i+1} k_{2i+1}} &= U_{i_{2i+1}j_{2i+1}}~, \qquad ~\, U_{1 k_{2i+1}}^* = U_{i'_{2i+1} j'_{2i+1}}^*~.
    \end{split}
\end{equation}
Contracting $b_{i+1}$ with itself corresponds to $\sigma(2i+1)=2i$. If we want to contract $k_{2i+1}$ with itself, we read off $\tau(2i+1)=2i+1$. From here, we already see we will need at least one transposition in $\sigma \tau^{-1}$ because $\sigma$ and $\tau$ act on $2i+1$ differently. Minimising the number of transpositions in $\sigma \tau^{-1},$ we see we thus have to take $\tau(2i) = 2i$ and $\sigma(2i) = 2i+1$. Then $\sigma \tau^{-1}$ has one extra transposition. After summing over the free variables with the exponential in front, this contributes a factor of $\frac{-W}{N} = -g$. 

If we were to not contract $k_{2i+1}$ with itself, we minimise the number of transpositions by taking $\tau(2i+1) = 2i$. By doing so, we have contracted $k_{2i+1}$ with $k_{2i}$. To introduce no further contractions, we thus take $\sigma, \tau(2i) = 2i+1$. This means we have introduced no further transpositions to $\sigma \tau^{-1}$ and this term contributes $N$ to the sum. 

We thus see that for $b_i = 2, \dots, N$, all leading order contributions have $\sigma = \prod_{i=1}^n \chi_i$ where $\chi_i$ is the transposition that contracts $b_i$ with itself:
\begin{equation}
    \chi_i = (2i-2,2i-1).
\end{equation}
Here $\chi_1 = (0,1) \equiv (1,2n)$. Including $\chi_i$ in $\tau$ is equivalent to contracting $k_{2i+1}$ and $k_{2i}$ with themselves. For each $\chi_i$, we can choose whether or not to include it in $\tau$. In total, we thus have $2^n$ contraction patterns that contribute to leading order to $\overline{ \Tr \left(\rho_{\mathsf{B}}^n\right)}$ when $b_i = 2, \dots, N$. We note that all of these contraction patterns couple different copies $\rho_{\mathsf{B}}$ to each other.
Summing these contributions, we find that in the large $N$ approximation,
\begin{equation}
   \overline{\Tr \left(\rho_{\mathsf{B}}^n\right)}\bigg \rvert_{b_i = 2, \dots, N} =   \sum_{\ell=0}^n \binom{n}{\ell} \overline{(-g)^\ell} = \overline{\left(1 - g\right)^n}.
\end{equation}
In total, there are $((2n)!)^2$ contraction patterns that contribute to the $n$'th R\'enyi entropy. However, because of the large $N$ limit, only $2^n+1$ of them contribute to leading order. In total, the $n$'th R\'enyi entropy of the environment for $n \geq 2$ in the large $N$ approximation is given by
\begin{equation}
    \overline{S(\rho_{\mathsf{B}})^{(n)}} = \frac{\log \left(\overline{\left(1 - g\right)^n} + \overline{g^n}\right)}{1-n}.
\end{equation}

\bibliography{references}
\bibliographystyle{JHEP}
\end{document}